\documentclass[11pt]{article}
\usepackage{amsmath,amssymb,graphicx,a4wide}

\begin{document}

\title{Aperiodic Crystals and Beyond}

\author{Uwe Grimm\\[2ex]
Department of Mathematics and Statistics, The Open University,\\
Walton Hall, Milton Keynes MK7 6AA, United Kingdom}

\maketitle

\begin{abstract}
  Crystals are paradigms of ordered structures. While order was once
  seen as synonymous with lattice periodic arrangements, the
  discoveries of incommensurate crystals and quasicrystals led to a
  more general perception of crystalline order, encompassing both
  periodic and aperiodic crystals. The current definition of crystals
  rests on their essentially point-like diffraction. Considering a
  number of recently investigated toy systems, with particular
  emphasis on non-crystalline ordered structures, the limits of the
  current definition are explored.
\end{abstract}

\section{What is order?}

The human brain is very skilled at detecting patterns and recognising
order in a structure, and ordered structures permeate cultural
achievements of human civilisations, be it in the arts, architecture
or music. The ability to detect and describe patterns is also at the
basis of all scientific enquiry; see Mumford \& Desolneux (2010) for
more on pattern theory. It may thus be surprising that a concept as
fundamental as \emph{order} does not have any well-defined precise
meaning, and that it appears to be rather challenging to come up with
a proper definition of what constitutes order in a structure. As a
consequence, there currently is no satisfactory measure to quantify
order in any given spatial structure.

There are two common approaches to tackle this issue. One is to employ
diffraction, which effectively measures two-point correlations in the
structure; see Cowley (1995) for background. For kinematic
diffraction, in the far-field approximation, the diffraction measure
is the Fourier transform of the autocorrelation (or Patterson)
function. Diffraction is the approach taken to characterise
crystalline materials. The current definition of a crystal, which is
based on its diffraction, derives from a definition which first
appeared in the terms of reference of the IUCr Commission on Aperiodic
Crystals, published in the 1991 report of the IUCr Executive Committee
(International Union of Crystallography, 1992, p.~928); in fact, it is
less general than what the commission proposed. The following quotes
the more specific definition given in Authier and Chapuis (2014), and
used in the IUCr Online Dictionary of Crystallography.

\begin{quote}
  A material is a crystal if it has \textbf{essentially} a sharp
  diffraction pattern. The word \textbf{essentially} means that most
  of the intensity of the diffraction is concentrated in relatively
  sharp \textbf{Bragg peaks}, besides the always present diffuse
  scattering. In all cases, the positions of the diffraction peaks can
  be expressed by
\begin{equation}\label{eq:crystal}
     \mathbf{H}\,=\sum_{i=1}^{n}h^{}_{i}\,\mathbf{a}_{i}^{*}\qquad 
     (n\ge 3).
\end{equation}
Here $\mathbf{a}_{i}^{*}$ and $h^{}_{i}$ are the basis vectors of the
reciprocal lattice and integer coefficients respectively and the
number $n$ is the minimum for which the positions of the peaks can be
described with integer coefficient $h^{}_{i}$.

The conventional crystals are a special class, though very large, for
which $n=3$.
\end{quote}

The prominent role of the word \emph{essentially} shows that this is a
humble definition, in the sense that it reflects our limited knowledge
of the structures we may potentially encounter in nature. The
interpretation given in the definition that `essentially' means that
most of the intensity is concentrated in Bragg peaks means that the
integrated contribution from the background must be weak compared to
the Bragg diffraction, which is rather arbitrary, as any Bragg
diffraction indicates order. By allowing the integer $n$ to be larger
than the three space dimensions we live in, aperiodic crystals are
included in this definition, and conventional (periodic) crystals have
become a special class (for which $n=3$). Note that $n$ is restricted
to be \emph{finite} here, so this particular form of the definition
excludes pure point diffractive systems with non-finitely generated
Fourier modules (over integer coefficients); the definition stipulates
that Bragg peaks in crystals can be indexed by a finite number of
integer coefficients. Note that the definition originally proposed in
1991 did not include this restriction (International Union of
Crystallography, 1992, p.~928).

Because the inverse problem of diffraction is inherently difficult
(Bombieri \& Taylor, 1986) and, in general, not unique (Patterson,
1944), we do not have a complete characterisation of structures that
show pure point diffraction (which means that the diffraction consists
of Bragg peaks only), even in the idealised case of an ideal, perfect
structure. Neither do we have a good understanding of what structures
with an \emph{essentially} pure point spectrum may look like.

The second approach, which is particularly suited to stochastic
systems, employs the \emph{entropy} of a structure. Entropy takes into
account the number of different local configurations of a system, and
how this number grows with the system size; normally you are looking
at an exponential growth with the system size, and any sub-exponential
growth corresponds to zero entropy. Clearly, entropy can distinguish
deterministic from random systems, and looking at different forms of
scaling behaviour makes it possible to differentiate, at least to some
extent, between different degrees of disorder. However, any
deterministic structure has zero entropy (as has any sufficiently
small deviation from it), so entropy is a rather crude measure of
order. 

In this article, the current state of knowledge of mathematical
diffraction of structures is summarised and discussed in relation with
our notion of crystalline order. The current article attempts to
present a broad overview only; for details on calculations and further
background, we refer to recent survey articles (Baake \& Grimm 2011a,
Baake \& Grimm 2012) and to the monograph by Baake \& Grimm (2013), as
well as to references therein.  Using a number of explicit example
structures with different types of diffraction spectrum, the range of
possibilities is explored, contributing to the ongoing discussion on
what order means in crystals and beyond (van Enter \& Mi\c{e}kisz
1992, Lifshitz 2003, 2007, 2011).

\section{Mathematical diffraction}

In 2014, we were celebrating the international year of
crystallography, and were looking back at a century of rapid
developments in crystallography since von Laue (Friedrich, Knipping \&
von Laue 1912, von Laue 1912) and father and son Bragg (Bragg \& Bragg
1913) first employed X-ray diffraction to analyse the atomic structure
of crystalline materials; see Authier (2013) for a historical
account. In the simplest setting, which is suitable in particular for
X-ray diffraction, it is sufficient to describe the kinematic
scattering of radiation by the sample, and consider the far-field
(Fraunhofer) limit for the outgoing radiation. The calculation of the
diffraction pattern of a given structure then becomes possible by
means of harmonic analysis, while the corresponding inverse problem of
determining a structure from its pattern of diffraction intensities
is, in general, difficult and non-unique, even in this simplified
setting. This section attempts to present a summary of mathematical
diffraction theory, highlighting the ideas and the flavour of the
approach without going into technicalities, while trying to explain
some of the technical terms by means of simple examples and familiar
notions; for mathematical details, the reader is referred to Baake \&
Grimm (2013). 

\subsection{What is a measure?}

A mathematically satisfactory approach to describe extended (infinite)
systems, such as ideal crystals, is provided by using \emph{measures}
to describe both the distribution of matter in the scattering medium
and the distribution of scattered radiation intensity in space. In
mathematics, measures are the natural concept to quantify spatial
distributions, and are related to the notion of integration. The
general approach to measures in mathematics is rather technical, but
there is a simpler way to think of measures (which is due to a result
called the Riesz-Markov representation theorem; see Reed \& Simon
(1980) for details). Indeed, it is possible to regard a measure as a
linear functional, which is a linear map that associates a number to
each function from an appropriate space. A familiar example is the
integral of a function, which is the example we start with.

A well-known and widely used measure in mathematics is \emph {Lebesgue
  measure}, which is commonly used in integration of functions on the
real numbers $\mathbb{R}$. We denote Lebesgue measure by the letter
$\lambda$. If $f\!: \; x\mapsto f(x)$ is a function on $\mathbb{R}$,
the Lebesgue measure of $f$ is
\[
    \lambda(f) \, = \, \int_{\mathbb{R}} f(x)\, \mathrm{d}\lambda(x)
               \, = \, \int_{\mathbb{R}} f(x)\, \mathrm{d}x \, ,
\]
where the usual shorthand $\mathrm{d}x$ is used for integration with
respect to Lebesgue measure. So Lebesgue measure associates to a
function $f$ a number, which is its integral.

The Lebesgue measure of a set $A\subset\mathbb{R}$, written as
$\lambda(A)$, is given by
\[
   \lambda(A) \, = \, \lambda(1_{A}) \, = \, 
                      \int_{\mathbb{R}} 1_{A}(x)\, \mathrm{d}x \, = \, 
                      \int_{A} \mathrm{d}x \, = \, \mathrm{vol}(A)\, ,
\]
where $1^{}_{A}$ is the characteristic function (or indicator
function) of $A$, which takes the value $1^{}_{A}(x)=1$ for all $x\in
A$, and $1^{}_{A}(x)=0$ otherwise. The Lebesgue measure of a set is
what we call the volume (as in this case we are in one dimension, the
volume is a length); for instance, the Lebesgue measure of the
interval $I=[a,b]$ with $b\ge a$ is $\lambda(I) = \lambda([a,b]) =
b-a$. Lebesgue measure is the unique translation-invariant measure on
$\mathbb{R}$ (meaning that $\lambda(A)=\lambda(A+t)$ for any
translation $t\in\mathbb{R}$) that assigns the volume $1$ to the unit
interval $[0,1]$.  Lebesgue measure on $\mathbb{R}$ generalises in the
familiar way to $d$-dimensional space $\mathbb{R}^d$, corresponding to
$d$-dimensional (multiple) integrals. For simplicity, we will mainly
consider the case $d=1$ in what follows.

Another well-known and important measure is the \emph{Dirac measure}
(or point measure) at a point $x\in\mathbb{R}$, denoted by
$\delta_{x}$. It describes a localised structure at a point $x$ in
space, with total measure $1$. This means that, if $f$ is a function
on $\mathbb{R}$, its point measure at $x$ is $\delta_{x}(f)=f(x)$. In
the physics literature, the point measure is often written like a
function $\delta(x)$ (which can be considered a generalised function
or \emph{distribution} obtained as a limit of functions, for instance
of a sequence of Gaussian functions of integral $1$, centred at the
origin, and with a decreasing width, which then become increasingly
sharper), with the suggestive notation
\[
    \delta_{x}(f) \, = \, \int_{\mathbb{R}} f(y)\, \delta(x-y)\, \mathrm{d}y
                  \, = \, f(x)\, .
\]
This notation can be used consistently as long as one remembers that
Dirac's $\delta$ is not a function in the usual sense. As above, one
can also define the point measure of a set $A\subset\mathbb{R}$, which
is $\delta_{x}(A)=\delta(1_{A})=1_{A}(x)$, so $\delta_{x}(A)=1$ if
$x\in A$ and $0$ otherwise.

\subsection{Dirac combs}

Point measures are often used to describe a set of localised
scatterers in space. Given a set of scatterers located at points in a
subset $\varLambda\subset\mathbb{R}$ (which we usually assume to be a
\emph{Delone set}, which means that it neither contains points that
are arbitrarily close to each other nor holes of arbitrary size), we
can associate a measure
\[
    \delta_{\varLambda}\, := \, \sum_{x\in\varLambda} \delta_{x}
\]
which we call the \emph{Dirac comb} (a term coined by de Bruijn
(1986); see also C\'{o}rdoba (1989)) of $\varLambda$. An example of
such a Dirac comb is
$\delta_{\mathbb{Z}}=\sum_{n\in\mathbb{Z}}\delta_{n}$, which is the
uniform Dirac comb on the integer lattice.

\begin{figure}
\centerline{\includegraphics[width=0.6\textwidth]{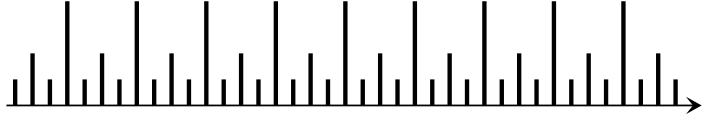}}
\caption{Schematic representation of the weighted (periodic) Dirac
  comb $\omega$ of Eq.~\eqref{eq:comb}. Point
  measures $a\,\delta_{x}$ are represented as columns at position $x$ of height
  proportional to their weight $a$.}
\label{fig:comb}
\end{figure}

By introducing scattering weights $w(x)$ at position $x\in\mathbb{R}$
(which in general can be complex numbers, but we will assume to be
real for the purpose of this exposition), we arrive at a \emph{weighted
Dirac comb}
\begin{equation}\label{eq:wdc}
    \omega \, = \, 
    w\, \delta_{\varLambda}\, = \, 
    \sum_{x\in\varLambda} w(x)\, \delta_{x}\, ,
\end{equation}
which can serve as a model representing a scattering medium containing
different types of scatterers. Any measure of this type, consisting of
a (weighted) sum of point measures, is called a \emph{pure point
  measure} (with respect to Lebesgue measure). It is possible to
include realistic scattering profiles by considering convolutions with
appropriate motives, so this approach is not as restrictive as it may
seem. A schematic representation of an example, the weighted (periodic)
Dirac comb
\begin{equation}\label{eq:comb}
\omega \, = \,  \delta_{\mathbb{Z}}+\tfrac{1}{2}\,\delta_{\mathbb{Z}+\frac{1}{2}} +
  \tfrac{1}{4}\,\delta_{\mathbb{Z}+\{\frac{1}{4},\frac{3}{4}\}}
\end{equation}
is shown in Figure~\ref{fig:comb}.

Attaching scattering profiles to a Dirac comb is one way to represent
a continuous scattering intensity in space. Of course, there is a more
direct approach if the scattering intensity is described by a
continuous distribution is space. If $\varrho$ is a such a continuous
(or, at least, locally integrable) function on $\mathbb{R}$, it
defines a measure $\mu$ on $\mathbb{R}$ via
\[
    \mu(f) \, = \, \int_{\mathbb{R}} f(x)\, \mathrm{d}\mu(x)
           \, = \, \int_{\mathbb{R}} f(x)\, \varrho(x)\, \mathrm{d}x \, .
\]
In this case $\varrho$ is called the \emph{density} of the measure
$\mu$. Any measure $\mu$ that can be written in this form is called an
\emph{absolutely continuous measure} (with respect to Lebesgue
measure).

The measures we are interested in are those which describe
distributions (of scatterers or scattering intensity) in space, and
one physical restriction we would like to impose is that any finite
region of space can only contain a finite total scattering strength or
finite intensity. The measures that satisfy this property are called
\emph{translation bounded} measures. A Dirac comb
$\delta_{\varLambda}$ of a Delone set $\varLambda\subset\mathbb{R}$ is
always translation bounded, because there can be only finitely many
points in any finite region of space, due to the minimum distance
between points. The same is true for a weighted Dirac comb provided
that the weight function $w(x)$ is bounded. An examples of a measure
that is not translation bounded would be a Dirac comb of a point set
with an accumulation point, for instance
$\sum_{n\in\mathbb{Z}\setminus\{0\}}\delta_{1/n}$. For this measure, any
interval containing the origin contains infinitely many point
measures, and thus has infinite measure.

\subsection{Lebesgue decomposition}

A central result in measure theory is Lebesgue's decomposition
theorem. It states that any measure can be written as a sum of three
components in a unique way (with respect to a reference measure, which
is our case will always be Lebesgue measure).  If $\mu$ is a measure
on $\mathbb{R}$, the three components are denoted as
\[
    \mu \, = \, \mu_{\mathsf{pp}} + \mu_{\mathsf{sc}} + \mu_{\mathsf{ac}}
\]
and are called the \emph{pure point} component $\mu_{\mathsf{pp}}$,
the \emph{singular continuous} component $\mu_{\mathsf{sc}}$ and the
\emph{absolutely continuous} component $\mu_{\mathsf{ac}}$. We have
met typical examples of pure point and absolutely continuous measures
above, so the obvious question is what a singular continuous measure
looks like. As it is defined, it is all that is `left' if you remove
the pure point part (consisting of a sum of weighted point measures)
and the absolutely continuous part (which is described by a locally
integrable density function) --- but this does not really help to gain
an understanding of what such a measure represents.  Singular
continuous measures are rather weird objects indeed; they do not give
weight to any single point in space (because otherwise it would have a
pure point component), but are concentrated on sets of vanishing
volume (because otherwise you could describe part of it by a density
function).

\begin{figure}
\centerline{\includegraphics[width=0.8\textwidth]{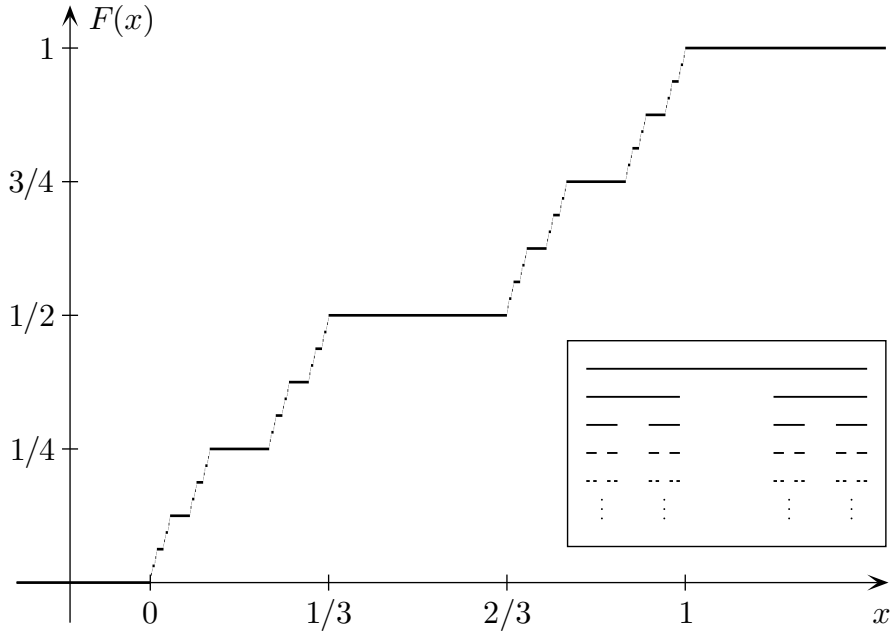}}
\caption{Sketch of the classic middle thirds Cantor set construction
  (inset) and the `Devil's staircase' distribution function $F$ of the
  corresponding singular continuous measure.}
  \label{fig:cantor}
\end{figure}

The best-known example of a singular continuous measure is provided by
the classic middle-thirds Cantor set; see Figure~\ref{fig:cantor}.
Starting from the unit interval, the Cantor set is constructed by
removing the middle third of it, then removing the middle thirds of
the two resulting intervals, and iterating this procedure \emph{ad
  infinitum}. The corresponding Cantor measure is constructed by
starting from the Lebesgue measure on the interval, so we have total
measure $1$, and at each step distributing the mass equally onto the
constituent intervals. In the limit, the total measure is thus still
$1$, but there is neither any isolated point that carries a finite
measure (since the measure of each interval in the $n$th step is
$2^{-n}$, so it vanishes in the limit) nor any interval of finite
length that is in the support of the measure (meaning that the measure
does not vanish on it). The measure constructed in this way is thus
singularly continuous, and can be described in terms of a distribution
function which is a `Devil's staircase'. This function is constant
almost everywhere, and displays a hierarchy of plateaux in its graph
(see Figure~\ref{fig:cantor}) which reflect the hierarchy of gaps
produced by the excision steps in the Cantor construction. This
function plays the role of the integrated density for the singular
continuous measure.

Lebesgue's decomposition provides a rigorous way to separate the
diffraction measure of a structure into its pure point (Bragg) part
and its singular and absolutely continuous components. However, using
this as the definition really only makes sense if one works with
infinite systems (because finite systems will always have absolutely
continuous diffraction). This is similar in spirit to the definition
of a phase transition in materials (as a discontinuity in a
thermodynamic potential), which again only applies in the
mathematically rigorous sense to an infinite system (because for
finite systems these potentials are smooth functions). Nevertheless,
these concepts have proved useful for applications to macroscopically
large (albeit finite) systems.

\subsection{Autocorrelation and diffraction}

A key quantity in diffraction theory is the \emph{autocorrelation},
which quantifies the two-point correlation of a structure. In
crystallography, this is often called the \emph{Patterson
  function}. If the material is described by a (real) density function
$\varrho$ on $\mathbb{R}$ (so we deal with an absolutely continuous
distribution), the autocorrelation is an absolutely continuous measure
whose density is the familiar convolution
\[
   P(x) \, = \, \int_{\mathbb{R}}\varrho(y)\, \varrho(y+x)\, \mathrm{d}x
        \, = \, \bigl(\varrho * \widetilde{\varrho}\bigr)(x)\, ,
\]
where $\widetilde{\varrho}$ is the function defined by
$\widetilde{\varrho}(x)=\varrho(-x)$. 

In the case that the material is described by a one-dimensional
weighted Dirac comb $\omega$ of the form given in Eq.~\eqref{eq:wdc}
(with real weight function $w(x)$), the autocorrelation is a
pure point measure 
\[
   \gamma \, = \, \sum_{z\in\varLambda-\varLambda} \eta(z)\, \delta_{z}\, ,
\]
with non-vanishing contributions only at distances $z$ in
the difference set $\varLambda-\varLambda=\{x-y\mid
x,y\in\varLambda\}$ (which you may interpret as the set of interatomic
distances). The point masses for interatomic distances $z$ are 
weighted by the \emph{autocorrelation coefficients} $\eta(z)$, 
which are given by
\[
    \eta(z) \,  =\,  \lim_{R\to\infty}\frac{1}{2R}
    \sum_{\substack{|x|\le R, x\in\varLambda \\ z-x\in\varLambda}} w(x)\,w(z-x)\, ,
\]
provided that these limits exist. Note that $2R$ is the length of the
interval $(-R,R)$, so the autocorrelation coefficient $\eta(z)$ is
precisely the volume-averaged two-point correlation for the
interatomic distance $z$.

Using the language of measures, these equations can be neatly summarised
as follows.  Given a (translation bounded) real measure $\omega$,
again for simplicity in one dimension, its \emph{autocorrelation
  measure} $\gamma$ is defined as
\begin{equation}\label{eq:def-gamma}
   \gamma \,   = \, \omega \circledast \widetilde{\omega} 
                 \, := \lim_{R\to\infty}
                \frac{\;\omega|^{}_{R} \ast \widetilde{\omega|^{}_{R}}\;}
                   {2R} ,
\end{equation}
provided the limit exists. Here, $\omega|^{}_{R}$ denotes the
restriction of $\omega$ to the interval $(-R,R)$, and
$\widetilde{\mu}$ is defined via $\widetilde{\mu}(g) =
\mu(\widetilde{g})$ with $\widetilde{g}(x)=\overline{g(-x)}$ as
above. Finally, $\ast$ denotes the standard \emph{convolution} of
measures, which is the appropriate generalisation of the convolution
of functions. The autocorrelation of $\omega$ is thus the
volume-averaged convolution $\circledast$ (also called the
\emph{Eberlein convolution}) of $\omega$ with its `flipped-over'
version $\widetilde{\omega}$, and thus picks out the two-point
correlations in the structure described by $\omega$. This approach to
mathematical diffraction was pioneered by Hof (1995).

The \emph{diffraction measure} is then the Fourier transform
$\widehat{\gamma}$ of the autocorrelation, so essentially it provides
a spectral analysis for the two-point correlations in the original
structure. It is a translation bounded, positive measure, which
quantifies how much of the kinematic scattering intensity reaches a
given volume in space. Lebesgue's decomposition
\[
   \widehat{\gamma} \; = \; 
            \widehat{\gamma}^{}_{\mathrm{pp} } +
            \widehat{\gamma}^{}_{\mathrm{sc} } +
            \widehat{\gamma}^{}_{\mathrm{ac} }
\]
into its pure point part (the Bragg peaks, of which there are at most
countably many), its absolutely continuous part (the diffuse
background scattering, described by a locally integrable density
function) and its singular continuous part (which encompasses anything
that remains) provides a mathematically rigorous definition of the
different types of diffraction. For the definition of a crystal cited
above, it is the pure point part $\widehat{\gamma}^{}_{\mathrm{pp} }$
that matters --- a crystal is a structure where this part represents
the majority of the diffracted intensity (there will always be some
continuous diffraction in practice), though this alone does not
guarantee that the \emph{positions} of Bragg peaks can be indexed by a
finite number of integers. Indeed, there are examples of systems that
are pure point diffractive, meaning that $\widehat{\gamma} =
\widehat{\gamma}^{}_{\mathrm{pp}}$, where this is not the case; we
shall meet an example below.

\section{Periodic crystals}

A conventional, periodic crystal is described as a lattice-periodic
structure, corresponding to an ideal infinite crystal without defects
or surfaces. A periodic crystal is characterised by its periods
(translations that keep the crystal invariant), which form a lattice
$\varGamma$ (because any linear combinations of periods are also
periods), and by the distribution of scatterers in a unit cell
(fundamental domain) of this lattice. Here, a lattice $\varGamma$ in
$d$-dimensional space\footnote{Because the generalisation to higher
  dimensions is straightforward, we give the results for the general
  case, although you can always think of cases with $d\le 3$; see also
  the examples below.} is the set of integer linear combinations of
$d$ linearly independent basis vectors, so it can be written in the
form
\[
    \varGamma \, :=\, \left\{\textstyle\sum_{i=1}^{d} a_{i}v_{i}\mid 
    a_{i}\in\mathbb{Z}\right\},
\]
where $v_{i}\in\mathbb{R}^{d}$, for $1\le i\le d$, are linearly
independent vectors in $\mathbb{R}^{d}$. Familiar examples are the
integer lattice $\mathbb{Z}$ in one dimension, the square lattice
$\mathbb{Z}^{2}$ in two dimensions and the simple (primitive) cubic
lattice $\mathbb{Z}^{3}$ in three dimensions.

If the scattering medium has a (periodic) crystal structure described
by a lattice $\varGamma$, it can always be represented as a measure
\begin{equation}\label{eq:cryst}
    \omega\, =\, \mu \ast \delta^{}_{\varGamma}\, ,
\end{equation}
where $\mu$ can be chosen as a finite measure which describes the
decoration of the fundamental domain, while the Dirac comb
$\delta^{}_{\varGamma}$ ensures lattice periodicity.

The corresponding autocorrelation measure is a $\varGamma$-periodic
measure that can be calculated from the appropriate generalisation of
Eq.~\eqref{eq:def-gamma} as
\begin{equation}\label{eq:crystauto}
     \gamma \, =\, \mathrm{dens} (\varGamma)\, (\mu \ast
     \widetilde{\mu}) \ast \delta^{}_{\varGamma}\, ,
\end{equation}
using $\widetilde{\delta^{}_{\varGamma}} =\delta^{}_{\varGamma}$ and
$\delta^{}_{\varGamma}\circledast\delta^{}_{\varGamma}= \mathrm{dens}
(\varGamma)\,\delta^{}_{\varGamma}$, where $\mathrm{dens}(\varGamma)$
denotes the density (per unit volume) of the lattice $\varGamma$. The
diffraction measure $\widehat{\gamma}$ is then given by\footnote{This
  follows from Eq.~\eqref{eq:crystauto} by an application of 
   Poisson's famous summation formula, which can be cast as
   $\widehat{\delta^{}_{\varGamma}} =
  \mathrm{dens} (\varGamma) \, \delta^{}_{\varGamma^{*}}$, where
  $\varGamma^{*}$ denotes the dual (or reciprocal) lattice of
  $\varGamma$; see Baake \& Grimm (2013) for details.}
\begin{equation}\label{eq:crystdiff}
     \widehat{\gamma} \, = \, \bigl(  \mathrm{dens} (\varGamma) \bigr)^{2}
     \, \big| \widehat{\mu} \big|^{2} \, \delta^{}_{\varGamma^{*}} \, .
\end{equation}
This provides the familiar result for periodic crystals: Any perfect
periodic crystal with lattice of periods $\varGamma$ shows pure point
diffraction with Bragg peaks located on the reciprocal
lattice\footnote{Note that we prefer to define the Fourier transform
  of a function $f$ as $\widehat{f}(k)=\int_{\mathbb{R}}
    \mathrm{e}^{2\pi\mathrm{i}kx}\,f(x)\,\mathrm{d}x$. Due to the
  factor $2\pi$ in the exponent, one avoids the appearance of such
  factors in the definition of the reciprocal lattice.}
$\varGamma^{*}$, and the intensity of the Bragg peak is determined by
the density of the crystal lattice $\varGamma$ and by the continuous
function $\big| \widehat{\mu} \big|^{2}$, which depends on the
distribution of scatterers in a fundamental domain of $\varGamma$. By
expressing the reciprocal lattice positions as linear combinations of
basis vectors of the reciprocal lattice $\varGamma^{*}$, this can be
cast in the form of Eq.~\eqref{eq:crystal} with $n=d$.

As a one-dimensional example, consider the weighted Dirac comb
$\omega$ of Eq.~\eqref{eq:comb} and Figure~\ref{fig:comb}. It can be
written as
\[
\omega \, = \,  \left(\delta_{0}+\tfrac{1}{4}\,\delta_{\frac{1}{4}}+
\tfrac{1}{2}\,\delta_{\frac{1}{2}}+\tfrac{1}{4}\,\delta_{\frac{3}{4}}\right) *
\delta_{\mathbb{Z}}\, ,
\]
so here $\mu = \delta_{0}+\frac{1}{4}\delta_{\frac{1}{4}}+\frac{1}{2}
\delta_{\frac{1}{2}}+\frac{1}{4}\delta_{\frac{3}{4}}$ describes the
four scatterers within the fundamental domain $[0,1)$ of the lattice
  $\varGamma=\mathbb{Z}$. Note that in this case we have
\[
\widetilde{\omega} \, = \,  
\left(\delta_{0}+\tfrac{1}{4}\,\delta_{-\frac{1}{4}}+\tfrac{1}{2}\,
\delta_{-\frac{1}{2}}+\tfrac{1}{4}\,\delta_{-\frac{3}{4}}\right) *
\delta_{\mathbb{Z}}\, = \, \omega\, ,
\]
due to the equivalence of positions differing by integers in the
periodic lattice and the symmetric distributions of scatterers in the
fundamental domain. Let us now calculate the autocorrelation and
diffraction of this comb. 

\begin{figure}
\centerline{\includegraphics[width=0.6\textwidth]{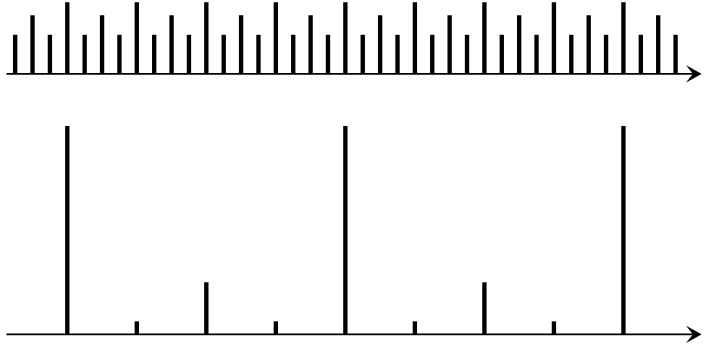}}
\caption{Schematic representation of the autocorrelation (top) and
diffraction (bottom) of the weighted Dirac
  comb $\omega$ of Eq.~\eqref{eq:comb} and Figure~\ref{fig:comb}.}
\label{fig:autodiff}
\end{figure}

Clearly, since all distances are multiples of $\frac{1}{4}$, the
autocorrelation in this case will have a similar form as the comb
$\omega$ itself, just with different coefficients, which are given by
summing up the products of the weights of scatterers with a given
separation.  To obtain these coefficients, one can compute the
convolution of $\omega$ with itself (or equivalently of $\mu$ with
itself) term by term, using the relation
$\delta_{x}*\delta_{y}=\delta_{x+y}$. This gives
\[
\gamma \, = \, 
\left(\tfrac{11}{8}\,\delta_{0}+\tfrac{3}{4}\,\delta_{\frac{1}{4}}+
\tfrac{9}{8}\,\delta_{\frac{1}{2}}+\tfrac{3}{4}\,\delta_{\frac{3}{4}}\right) * 
\delta_{\mathbb{Z}}\, .
\]
For instance, the coefficient $\frac{11}{8}=
1^2+(\frac{1}{4})^2+(\frac{1}{2})^2+(\frac{1}{4})^2$ of $\delta_{0}$
comes from adding up the contributions to integer distances. A
schematic presentation of the autocorrelation $\gamma$ is shown in
Figure~\ref{fig:autodiff}.

The corresponding diffraction measure $\widehat{\gamma}$ 
is obtained by taking the Fourier transform, 
using that $\widehat{a\delta_{x}}=a\mathrm{e}^{2\pi\mathrm{i}kx}$. This gives
\[
\begin{split}
\widehat{\gamma} \, &= \, 
\left(\tfrac{11}{8}+\tfrac{9}{8}(-1)^{k}+
\tfrac{3}{2}\cos(\tfrac{\pi k}{2})\right) *
\delta_{\mathbb{Z}}\\
&=\, 4\,\delta_{4\mathbb{Z}} + \delta_{4\mathbb{Z}+2} + 
\tfrac{1}{4}\,\delta_{4\mathbb{Z}+\{1,3\}}\, .
\end{split}
\]
The diffraction pattern is thus periodic, but with period $4$ (due to
the smallest distance between scatterers being $\frac{1}{4}$). A
schematic picture of the diffraction pattern is shown in
Figure~\ref{fig:autodiff}.

\begin{figure}
\centerline{\includegraphics[width=\textwidth]{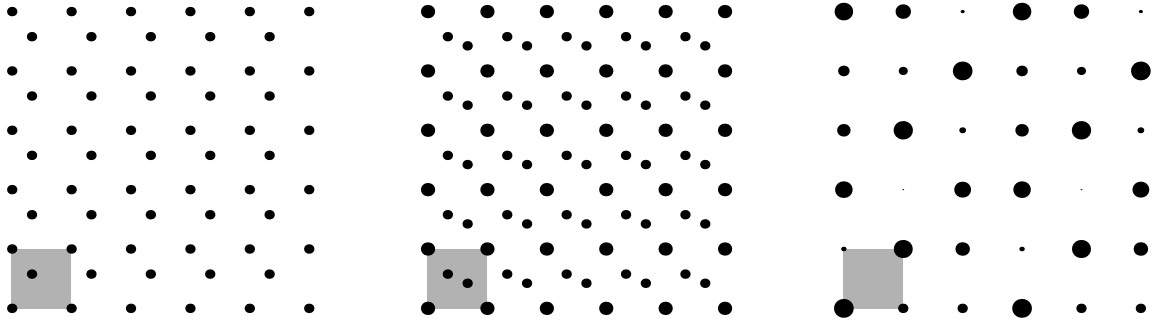}}
\caption{The left panel shows a schematic representation of the
  two-dimensional toy crystal discussed in the text, with two
  scatterers of equal strength at positions $(0,0)$ and
  $(a,b)=(\frac{1}{3},\frac{1}{\sqrt{3}})$ of the fundamental domain
  (indicated by shading).  The corresponding autocorrelation $\gamma$
  is shown in the central panel, while the corresponding diffraction
  measure $\widehat{\gamma}$ is displayed in the right panel. Here, a
  point measure is represented by a dot that is centred at the
  position of the peak and has an area proportional to the weight of
  the point measure. The irrational position in the vertical direction
  leads to an incommensurate modulation of the peak intensities in
  this direction, which the diffraction is periodic with period $3$ in
  the horizontal direction.}
\label{fig:twod}
\end{figure}

As a second example, consider a two-dimensional crystal with lattice
of periods $\varGamma=\mathbb{Z}^{2}$, with two scatterers (of unit
scattering strength) per unit cell, one placed at lattice points and
the other at an arbitrary position $(a,b)\in [0,1)^2$. The
  corresponding point set is $\varLambda=\mathbb{Z}^{2}\cup
  \bigl(\mathbb{Z}^2+(a,b)\bigr)$, and the Dirac comb can be written
  as $\omega=\varrho * \delta_{\mathbb{Z}^2} =
  (\delta_{0,0}+\delta_{a,b}) * \delta_{\mathbb{Z}^2}$. Its
  autocorrelation is
\[
\begin{split}
   \gamma & \, =\, 
   \bigl(\varrho * \widetilde{\varrho}\bigr)*\delta^{}_{\mathbb{Z}^2}\\
    & \, = \, 
   (\delta_{(0,0)}+\delta_{(a,b)}) *
   (\delta_{(0,0)}+\delta_{-(a,b)}) * \delta^{}_{\mathbb{Z}^2}\\ 
   & \, = \, \bigl(2\,\delta_{(0,0)} + \delta_{(a,b)}+\delta_{-(a,b)}\bigr) 
   * \delta^{}_{\mathbb{Z}^2}  .
\end{split}
\]
The corresponding diffraction measure is then 
\[
\begin{split}
  \widehat{\gamma} &\, =\, \lvert\widehat{\varrho}\,\rvert^{2}(k,\ell) 
  \,\delta^{}_{\mathbb{Z}^{2}} \\ 
  &\, = \, 
  \bigl(2 + 2\,\mathrm{Re}(\mathrm{e}^{-2\pi \mathrm{i} (ka+\ell b)})\bigr)\,
   \delta^{}_{\mathbb{Z}^2}\\
   &\, = \, \left(2+2\cos\bigl((2 \pi  (k a + \ell b)\bigr)\right)\,
   \delta^{}_{\mathbb{Z}^2}\\
   &\, = \, \left(2\cos\bigl(\pi  (k a + \ell b)\bigr)\right)^{2}\,
   \delta^{}_{\mathbb{Z}^2}
\end{split}
\]
for $k,\ell\in\mathbb{Z}$. Note that, while the diffraction measure is
supported on $\mathbb{Z}^{2}$ as expected (as $\mathbb{Z}^{2}$ is
self-dual), it need not have any non-trivial period.  In fact,
$\widehat{\gamma}$ is periodic in one or two directions precisely if
one or both coordinates $a$ and $b$ are rational, respectively; an
example with one periodic direction is shown in Figure~\ref{fig:twod}.
Although the positions of Bragg spots for a lattice-periodic structure
are again lattice-periodic, in general the intensities will not
respect the periodicity of the dual lattice.

\section{Aperiodic crystals}

The arguably best understood class of aperiodic structures are cut and
project sets, also called \emph{model sets}.  Model sets can be viewed
as a natural generalisation of the notion of quasiperiodic functions,
which goes back to Harald Bohr (1947), and were first introduced by Yves Meyer
(1972) in the context of harmonic analysis. The basic idea of the
construction is to obtain an aperiodic structure as a suitable `slice'
of a higher-dimensional periodic lattice, which is then projected onto
a suitable space of the desired dimension. For simplicity, we mainly
consider the case where the higher-dimensional space is a Euclidean
space of the form $\mathbb{R}^{d+m}$, with $\mathbb{R}^{d}$ being the
physical space (sometimes also called the direct or the parallel
space) that hosts the aperiodic structure (so $1\le d\le 3$ for
physically relevant cases), and $\mathbb{R}^{m}$ the internal (or
perpendicular) space which is used in the construction.

\begin{figure}
\centerline{\includegraphics[width=0.9\textwidth]{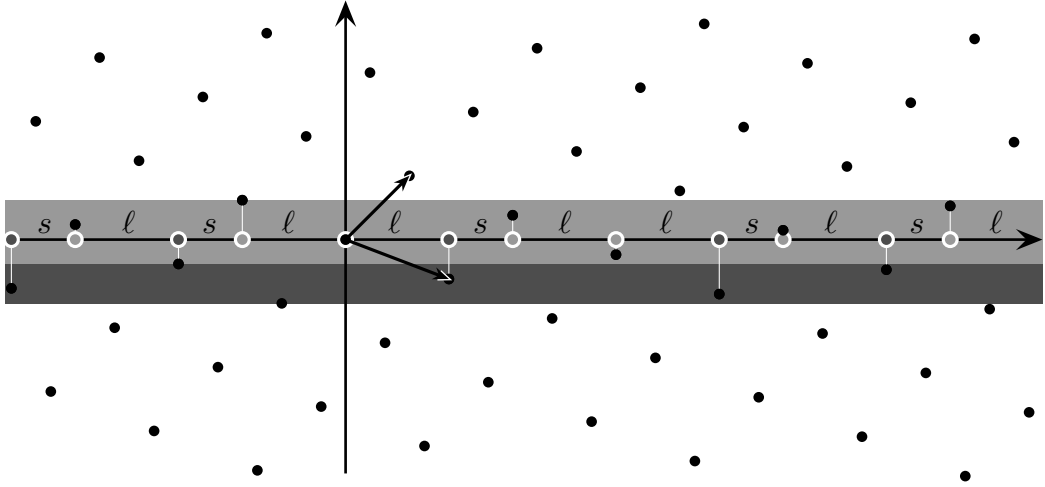}}
  \caption{Schematic representation of a natural projection approach
    for the Fibonacci chain from the planar lattice spanned by the
    vectors $(1,1)$ and $(\tau,1-\tau)$.}
\label{fig:fiboproj}
\end{figure}

Let us start with an example, where $d=m=1$. In this case, we are
projecting a one-dimensional aperiodic structure from a
two-dimensional (periodic) lattice. In this example, the lattice
$\mathcal{L}$ is given by all integer linear combinations of two basis
vectors, which we choose as $(1,1)$ and $(\tau,1-\tau)$, where
$\tau=(1+\sqrt{5})/2$ is the golden ratio, so we have
\[
   \mathcal{L} \, = \, 
   \mathbb{Z}\, (1,1) + \mathbb{Z}\, (\tau,1-\tau) \, = \,
   \bigl\{ m \, (1,1)+n\, (\tau,1-\tau) \;\big| \; 
   m,n\in\mathbb{Z}\bigr\} \, . 
\]
The lattice points are shown as black dots in
Figure~\ref{fig:fiboproj}. The lattice is oriented such that the
horizontal space along the $(1,0)$ direction is the physical space and
the vertical direction along $(0,1)$ corresponds to the internal
space. We call the projection to the physical space $\pi$, and the
projection to the internal space $\pi_{\mathrm{int}}$. The projections
of all lattice points $L=\pi(\mathcal{L})$ to physical space and
$L^{\star}=\pi_{\mathrm{int}}(\mathcal{L})$ to internal
space\footnote{Note the difference between the star symbol $\star$
  used here and the $*$ used for the dual or reciprocal lattice.} are
both dense in their corresponding one-dimensional spaces.  The set
$L$ is explicitly given by $L=\mathbb{Z}[\tau]=\{m+n\tau\mid
m,n\in\mathbb{Z}\}$, so all integer combinations of multiples of $1$
and $\tau$, which is dense because $\tau$ is irrational, and $L^{\star}$
has the same form. Note that the projections are one-to-one in both
directions. In particular, any point in $L$ corresponds to a uniquely
defined point in $\mathcal{L}$. In fact, $\pi^{-1}(x)=(x,x^{\star})$,
where the $\star$-map is defined by mapping $1$ to $1$ and $\tau$ to
$1-\tau$ (which corresponds to the `algebraic conjugation'
$\sqrt{5}\mapsto -\sqrt{5}$), so $(m+n\tau)^{\star}=m+n(1-\tau)=m+n-n\tau$
for all $m,n\in\mathbb{Z}$.

The final ingredient that we require is a `window' $W$ in the internal
space, which we choose to be the half-open interval $W=(-1,\tau-1]$.
Shifting it along the physical space sweeps out the shaded horizontal
strip in Figure~\ref{fig:fiboproj}. The lattice points that fall
within this strip produce the set $\{x\in\mathcal{L}\mid
\pi^{}_{\mathrm{int}}(x) \in W\}$, and their projection onto the
physical space thus $\varLambda=\{\pi(x)\mid x\in\mathcal{L}\text{ and
} \pi^{}_{\mathrm{int}}(x) \in W\}$. Using $\pi(\mathcal{L})=L$ and
the $\star$-map, this point set can equivalently be written as
\begin{equation}\label{eq:fiboms}
   \varLambda \, = \, \{ x\in L \mid x^{\star}\in W\}\, .
\end{equation}
Sets of this form are called cut and project sets or model sets.
The condition that $x^{\star}\in W$ selects a discrete subset of the dense
set point set $L$, in fact, a very special discrete subset where
points are separated either by intervals of length $1$ (for short
intervals $s$) or by intervals of length $\tau$ (for long intervals
$\ell$). As it turns out, this projection yields the famous Fibonacci
sequence $\dots \ell s \ell\ell s \ell s \ell\dots $ of long ($\ell$)
and short ($s$) intervals, which can be generated by the two-letter
substitution rule $\ell\mapsto \ell s$, $s\mapsto \ell$. In
particular, dividing the window into two parts as follows
\[
    W_{s} \, = \, (\tau-2,\tau-1]\quad\text{and}\quad
    W_{\ell} \, = \, (-1,\tau-2]
\]
shows that the sets of left endpoints of short or long intervals are
given by the projection of lattice points that fall within the
corresponding sub-strip, so
\[
   \varLambda_{s} \, = \, \{ x\in L \mid x^{\star}\in W_{s}\}\quad\text{and}\quad
   \varLambda_{\ell} \, = \, \{ x\in L \mid x^{\star}\in W_{\ell}\}\, ,
\]
with $\varLambda=\varLambda_{s}\cup\varLambda_{\ell}$. Hence the set of
left endpoints of short or of long intervals separately are model sets
with windows $W_{s}$ and $W_{\ell}$, respectively, while the set of all
left interval endpoints is a model set with window
$W=W_{s}\cup W_{\ell}$; compare Figure~\ref{fig:fiboproj}.

As mentioned above, this construction can be generalised to physical
and internal spaces of any dimension.  The general \emph{cut and
  project scheme} (CPS) for Euclidean model sets can be summarised in
the following diagramme
\begin{equation}\label{eq:cps}
\renewcommand{\arraystretch}{1.2}\begin{array}{r@{}ccccc@{}l}
   & \mathbb{R}^{d} & \xleftarrow{\,\;\;\pi\;\;\,} 
         & \mathbb{R}^{d} \times \, \mathbb{R}^{m}\!  & 
        \xrightarrow{\;\pi^{}_{\mathrm{int}\;}} & \mathbb{R}^{m} & \\
   & \cup & & \cup & & \cup & \hspace*{-2ex} 
   \raisebox{1pt}{\text{\footnotesize dense}} \\
   & \pi(\mathcal{L}) & \xleftarrow{\; 1-1 \;} & \mathcal{L} & 
   \xrightarrow{\; \hphantom{1-1} \;} & 
       \pi^{}_{\mathrm{int}}(\mathcal{L}) & \\
   & \| & & & & \| & \\
   &  L & \multicolumn{3}{c}{\xrightarrow{\qquad\quad\quad \;\;
       \;\star\; \;\; \quad\quad\qquad}} 
       &   {L_{}}^{\star} & \\
\end{array}\renewcommand{\arraystretch}{1}
\end{equation}
Here, $\mathcal{L}\subset \mathbb{R}^{d+m}$ is a lattice in the
$(d+m)$-dimensional space
$\mathbb{R}^{d}\times\mathbb{R}^{m}=\mathbb{R}^{d+m}$, and $\pi$ and
$\pi^{}_{\mathrm{int}}$ denote the natural projections from this space
onto the physical and internal spaces $\mathbb{R}^{d}$ and
$\mathbb{R}^{m}$, respectively. We assume that the point set
$L=\pi(\mathcal{L})\subset\mathbb{R}^{d}$, which is the projection of
the lattice points into the physical space, is a bijective image of
$\mathcal{L}$, which means that no two lattice points in $\mathcal{L}$
project onto the same point in $L$. In other words, each point in $L$
can be `lifted' to a unique lattice point in $\mathcal{L}$, and the
inverse map $\pi^{-1}$ is well-defined on all elements of $L$.  This
ensures that the star-map $\star\!:\; x\mapsto x^{\star}$ is
well-defined on $L$, so each point in $L$ has a unique associate in
internal space; see Moody (2000) for details. Finally, we assume
that the corresponding set
$L^{\star}=\pi^{}_{\mathrm{int}}(\mathcal{L})\subset\mathbb{R}^{m}$ in
internal space is dense.

Given a CPS, the second ingredient in the definition of a cut and
project set is the \emph{window} (sometimes also called an acceptance
domain) $W\subset\mathbb{R}^{m}$, which is assumed to be a
sufficiently nice subset of $\mathbb{R}^{m}$ (technically, a
relatively compact subset with non-empty interior). A cut and project
set is then defined by selecting all points $x$ in the projected
lattice $L$ whose companion $x^{\star}$ in internal space falls inside
the window $W$. Expressed as an equation, this means that any set of
the form
\begin{equation}\label{eq:ms}
    \varLambda \, = \,
    \bigl\{  x\in L \mid  x^{\star} \in W \bigr\} ,
\end{equation}
or indeed any translate of such a set, is what we call a \emph{model
  set}. The technical conditions on the window $W$ ensure that
$\varLambda$ is a Meyer set (Meyer 1972, Moody 2000), which means that
the difference set
\[
  \varLambda-\varLambda\, :=\, \{x-y\mid x,y\in\varLambda\} 
\]
is uniformly discrete (so different distances between point in the
structure differ by at least a fixed amount) and that the set
$\varLambda$ is relatively dense (which means that there are no
arbitrarily large `holes' in the point set). If the boundary $\partial
W$ of the window $W$ is nice in the sense that it has zero volume (in
the sense of Lebesgue measure), we refer to $\varLambda$ as a
\emph{regular model set}.  The setting of Eq.~\eqref{eq:cps} can be
generalised further to allow for the internal space to be a locally
compact Abelian group (Meyer 1972, Moody 2000, Schlottmann 2000).

\begin{figure}
\centerline{\includegraphics[width=0.9\textwidth]{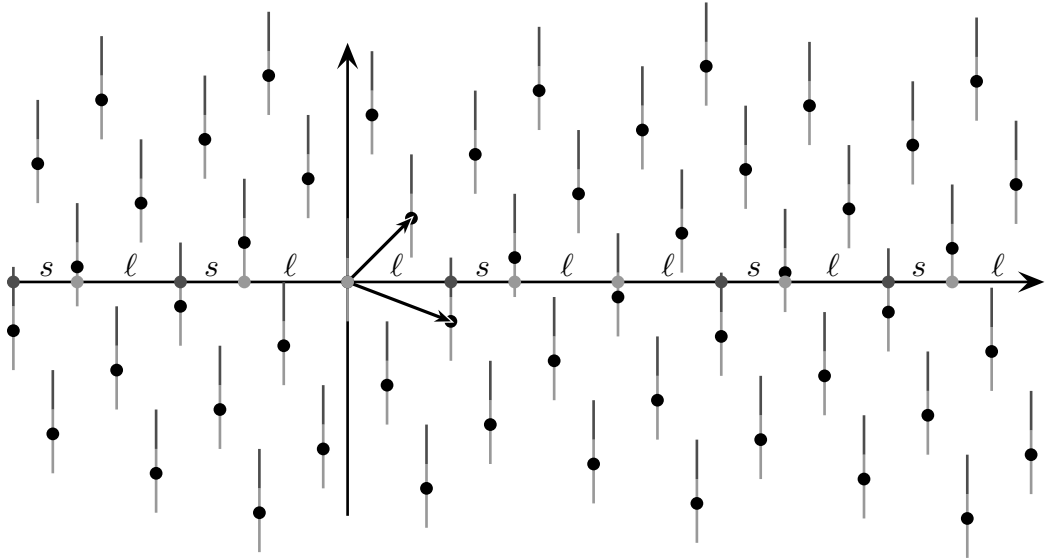}}
\caption{Equivalent description of the Fibonacci chain in terms of 
`targets', often referred to as `occupation domains' or `atomic surfaces'.}
\label{fig:fibotarget}
\end{figure}

It is worth mentioning that there are various equivalent ways of
interpreting the cut and project construction. One commonly used
approach attaches an inverted copy of the window as a `target' to each
lattice point, and projected points are then obtained as the
intersection of these targets with the physical space; see
Figure~\ref{fig:fibotarget} for an illustration of the Fibonacci case.
Albeit equivalent, this description offers a simpler way of
interpreting experimental data, and is therefore the preferred
presentation of the cut and project approach in experimental research
papers, where it is often referred to as the \emph{section
  method}. Apart from providing an intuitive meaning for the targets
as `atomic surfaces', this approach allows for additional variation
(by deformations of the targets) that can be exploited, for instance
in the description of modulated phases. For further variants of the
cut and project method, the reader is referred to Chapter~7.5 in Baake
\& Grimm (2013).

Familiar examples of model sets are some one-dimensional
substitution-based structures such as the Fibonacci chain discussed
above, and some of its generalisations. Planar examples include the
Penrose tiling and the T\"{u}bingen triangle tiling with decagonal
symmetry, the Ammann-Beenker tiling with octagonal symmetry and the
shield tiling with dodecagonal symmetry, which can all be obtained by
projection from four-dimensional lattices. Structure models of
icosahedral quasicrystals usually employ model sets based on the
(primitive, face-centred or body-centred) hypercubic lattice in six
dimensions. These examples have direct application to the
crystallography of quasicrystals, and serve as models for the
structure of decagonal, octagonal, dodecagonal and icosahedral
quasicrystals, respectively; compare Steurer \& Deloudi (2009) for
details. Realisations of model sets with other symmetries, such as
planar sevenfold symmetry for instance, have as yet not been observed
in nature, and the same is true for model sets where the internal
space is not Euclidean. Nevertheless, such systems share many
properties with the familiar quasicrystalline cases, and should not be
excluded \emph{a priori}. Even if it may prove impossible to realise
such structures in self-assembled systems, we may be able to produce
these in purpose-made manufactured structures at various length
scales, from macroscopic to nanometre and atomic scales.

Arguably the most important result in the theory of model sets, in the
context of mathematical diffraction theory, is the proof that regular
model sets have \emph{pure point diffraction}. Three different
approaches have been used to prove this statement. The first proof
using methods of dynamical systems theory was completed by Schlottmann
(2000), employing an argument by Dworkin (1993) and the mathematical
diffraction approach of Hof (1995); see also Lenz \& Strungaru (2009)
for further developments. An alternative approach employs the theory
of almost periodic measures (Baake \& Moody 2004, Moody \& Strungaru
2004, Strungaru 2005). Following a suggestion by Lagarias, Baake \&
Grimm (2013) present a proof based on Poisson's summation formula for
the embedding lattice in conjunction with Weyl's lemma on uniform
distribution, which exploits the uniform distribution of projected
lattice points in internal space. Although it has not been developed
into a proof, there is also a somewhat complementary approach based on
an average periodic structure; we refer to the recent review by Wolny,
Kozakowski, Kuczera, Strzalka \& Wnek, A. (2011) and references
therein for details.

\begin{figure}[t]
\centerline{\includegraphics[width=0.9\textwidth]{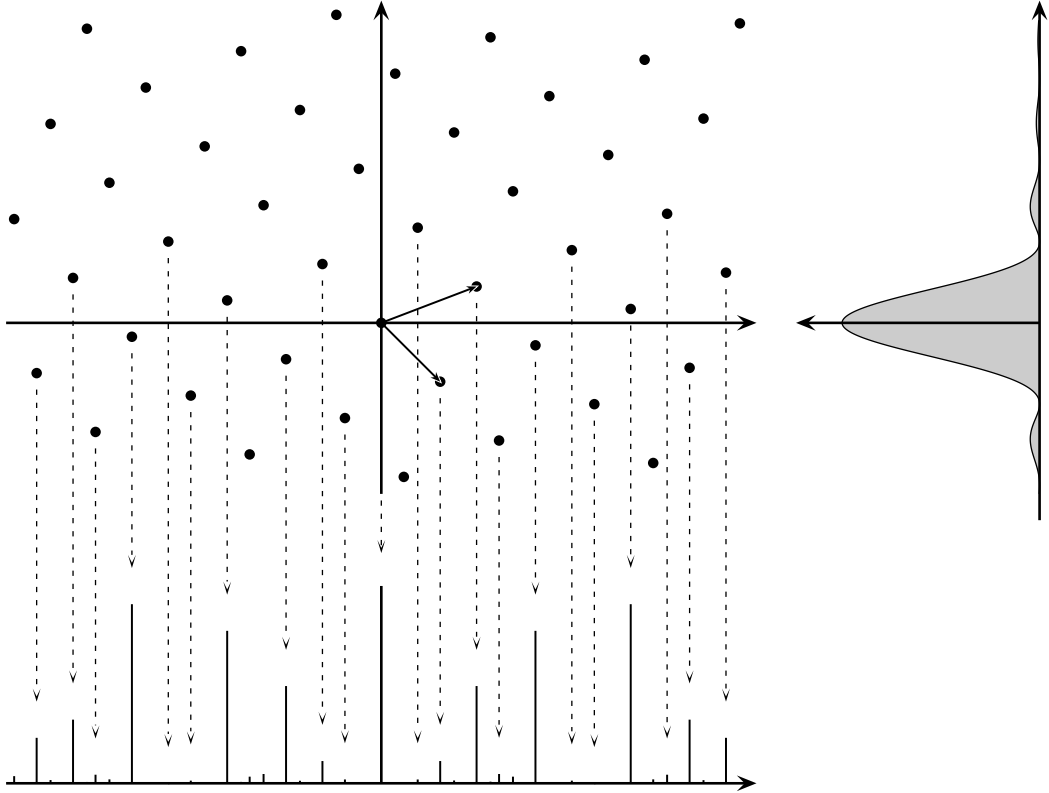}}
\caption{Sketch of the projection of the dual lattice points giving
  rise to Bragg peaks in the diffraction pattern for the Fibonacci
  point set $\varLambda$, with scatterers of unit weight at all
  points. The function displayed on the right-hand side is the
  intensity function $\lvert A(k)\rvert^{2}$.  The Bragg peak at $0$
  has height $(\mathrm{dens} (\varLambda))^{2} = (\tau+1)/5\approx
  0.5206$, and the entire pattern (once all reflections are included)
  is reflection symmetric.}
  \label{fig:fibodiff}
\end{figure}

Essentially, the pure point diffraction of a model set is a
consequence of the underlying higher-dimensional lattice periodicity.
Let us first discuss the example of the Fibonacci model set
$\varLambda$ of Eq.~\eqref{eq:fiboms}; compare also
Figure~\ref{fig:fiboproj}. The pure point diffraction pattern is
obtained again as a projection to physical space, but this time of the
\emph{dual} (or reciprocal) higher-dimensional lattice
$\mathcal{L}^{*}$. In our case, this is the lattice generated by the
dual basis vectors $\frac{2\tau-1}{5}(\tau-1,\tau)$ and
$\frac{2\tau-1}{5}(1,-1)$. The corresponding Fourier module is then
\[
  L^{\circledast} \, =\, \pi (\mathcal{L}^{*}) \, = \,
     \frac{1}{\sqrt{5}}\, \mathbb{Z}[\tau] \, ,
\]
where $\mathbb{Z}[\tau]=\{m+n\tau\mid m,n\in\mathbb{Z}\}$ as above.
The determines the positions of Bragg peaks, but what about their
intensities? It turns out that the intensity is a function of the
distance of the projected lattice point from the physical space, and
roughly the larger the internal coordinate the smaller the intensity. 
The function in question is the absolute square of the Fourier transform
of the window function (the characteristic function of the window), which is the
function that takes the value $1$ on the window and $0$ otherwise. Its Fourier
transform is  
\begin{equation}
   A(k) \, = \, \mathrm{e}^{\pi \mathrm{i} k^{\star} (\tau-2)}\,\frac{\tau + 2}{5}\, 
   \mathrm{sinc}(\pi \tau k^{\star}) \, , 
\end{equation}
where $\mathrm{sinc}(x)=\sin(x)/x$, and $k^{\star}$ is the image of
$k$ under the $\star$-map introduced above. A sketch of the
diffraction pattern is shown in Figure~\ref{fig:fibodiff}.

Let us now return to the general result.
For a regular model set $\varLambda$ with Dirac comb
$\delta^{}_{\varLambda}$, the diffraction measure $\widehat{\gamma}$
can be written as
\begin{equation}\label{eq:modeldiff}
    \widehat{\gamma}\,  = \sum_{k\in L{}_{}^{\circledast}}
         \lvert A(k) \rvert^{2}\, \delta_{k}\, ,
\end{equation}
where $L^{\circledast} = \pi (\mathcal{L}^{*})$, the projection of the
higher-dimensional dual lattice, is the corresponding Fourier module
on which the pure point diffraction is supported. For a Euclidean
model set with the CPS \eqref{eq:cps}, $\mathcal{L}$ is a lattice in a
Euclidean space $\mathbb{R}^{d+m}$, and the Fourier module
$L^{\circledast}$ is thus finitely generated, with rank $d+m$. By
choosing appropriate generating vectors, the pure point diffraction of
Eq.~\eqref{eq:modeldiff} can thus be recast in the form of
Eq.~\eqref{eq:crystal} with $n=d+m$. However, in the general
situation, where the internal space can be any locally compact Abelian
group, this is not necessarily the case, as the Fourier module
$L^{\circledast} = \pi (\mathcal{L}^{*})$ does not have to be finitely
generated. Note that the latter case is not covered by the definition
of a crystal cited above, while it does include any model set based on
a Euclidean CPS.

The diffraction amplitudes $A(k)$ are obtained by the Fourier
transform of the characteristic function $1^{}_{W}$ of the window $W$,
\begin{equation}\label{eq:modelamp}
   A(k) \, = \, 
  \frac{\mathrm{dens} (\varLambda)}{\mathrm{vol} (W)}
  \, \widehat{1^{}_{W}} (-k^{\star})\, .
\end{equation}
According to Eq.~\eqref{eq:modeldiff}, it is the absolute square of
these amplitudes that determine the intensity of a Bragg peak as
position $k\in L^{\circledast}$, with $k^{\star}$ denoting the
corresponding point in internal space. Eq.~\eqref{eq:modelamp} gives
the result for Euclidean model sets, the only difference for the
general case is that the volume (with respect to Lebesgue measure in
Euclidean space) is replaced by the suitable invariant measure (Haar
measure) on the locally compact Abelian group.

\section{Order beyond crystals}

The current definition of crystals thus covers conventional periodic
crystals, incommensurate crystals and quasicrystals, and hence all
currently known realisations of perfectly ordered structures in
nature. While the question asked by Bombieri \& Taylor (1986) has not
yet been satisfactorily answered, it is clear that pure point
diffraction is a severe constraint on the possible structure (Baake,
Lenz \& Richard 1997), and recent results by Lenz \& Moody (2009,
2011) indicate that model sets play a major role in a potential
abstract parametrisation of the inverse problem. However, there are
clearly well-ordered structures that do not possess this property, and
this section will discuss a few characteristic examples. But first, we
start with an example of a pure point diffractive system with
non-finitely generated Fourier module, which thus possesses a
diffraction pattern where Bragg peaks cannot be indexed by a finite
number of integers.

\subsection{Pure point diffraction with non-finitely generated support}

Well-known examples of systems with non-Euclidean internal spaces
are limit-periodic structures. Let us explain this with the arguably
simplest example, based on the \emph{period doubling} substitution
rule $\varrho\!: 1\mapsto 10, 0\mapsto 11$, on the two-letter alphabet
$\{0,1\}$. Any bi-infinite word\footnote{Here and below, the notation
  $\mathcal{A}^{\mathbb{Z}}$ denotes the set of bi-infinite sequences
  $(\ldots,a^{}_{-2},a^{}_{-1},a^{}_{0},a^{}_{1},a^{}_{2},\ldots)$
  with letters $a_{i}$, $i\in\mathbb{Z}$, chosen from a finite
  alphabet $\mathcal{A}$.}  $w\in\{0,1\}^{\mathbb{Z}}$ that satisfies
the fixed point property $\varrho^{2}(w)=w$ is specified completely by
$w(2n)=1$, $w(4n+1)=0$ and $w(4n+3)=w(n)$ for $n\in\mathbb{Z}$, while
either letter can be chosen at position $n=-1$. The two possible
choices lead to two locally indistinguishable sequences (which means
that any finite subsequence of one occurs in the other), and hence
define the same system.

\begin{figure}[t]
\centerline{\includegraphics[width=0.8\textwidth]{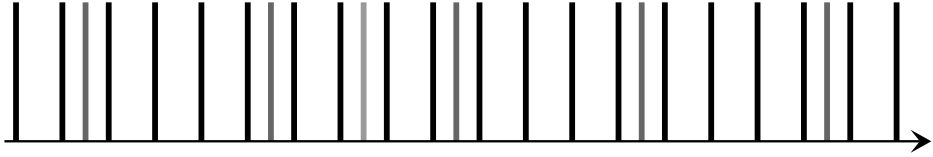}}
\caption{Schematic representation of the Dirac comb
  $\delta_{\varLambda}$ of the period doubling point set of
  Eq.~\eqref{eq:toep}. All point measures have the same mass. The
  different shading highlights the Toeplitz structure, with point
  masses at even integers shown in black, point masses on
  $8\mathbb{Z}+3$ in dark grey and a single point mass in
  $32\mathbb{Z}+15$ in lighter grey.}
\label{fig:toep}
\end{figure}

The word $w$ possesses a Toeplitz structure consisting of a hierarchy
of scaled and shifted copies of $\mathbb{Z}$ which carry the same
letter. Defining the point set 
\begin{equation}\label{eq:toep}
  \varLambda\, =\, \{n\in\mathbb{Z}\mid w(n)=1\}\subset\mathbb{Z}
\end{equation}
 of the positions of the letter $1$ in $w$,
it is clear from the relations above that
$2\mathbb{Z}\subset\varLambda$, as all letters at even positions are
$1$.  But then, due to $w(4n+3)=w(n)$, so are all letters
$w(8n+3)=w(2n)=1$, so $8\mathbb{Z}+3\subset\varLambda$, and
inductively one recognises that $2\cdot 4^{\ell}\mathbb{Z} +
(4^{\ell}-1) \subset\varLambda$ for all integer $\ell\ge 0$. In fact,
this hierarchy of scaled integer lattices describes the complete set,
and we obtain the following representation for the set as a union
(Baake, Moody \& Schlottman 1998, Baake \& Moody 2004, Baake \& Grimm
2013)
\begin{equation}\label{eq:pdps}
     \varLambda \, = \, 2\,\mathbb{Z}\cup (8\,\mathbb{Z}+3) \cup 
     (32\,\mathbb{Z}+15) \cup \dots 
      \, = \, 
      \bigcup_{\ell\ge 0} \bigl(( 2\cdot 
      4^{\ell}\,\mathbb{Z} + (4^{\ell}-1)\bigr) 
\end{equation}
of scaled (and shifted) lattices. Note that this result is for the
case where we choose $w(-1)=0$ (otherwise $-1$ has to be added to the
right-hand side). A schematic representation of the corresponding
Dirac comb $\delta_{\varLambda}$ is shown in Figure~\ref{fig:toep}.

\begin{figure}[b]
\centerline{\includegraphics[width=0.8\textwidth]{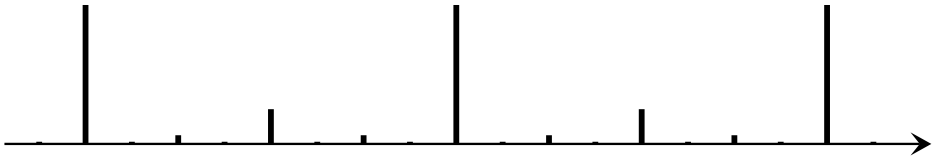}}
\caption{Schematic representation of the diffraction intensity pattern
  of the Dirac comb $\delta_{\varLambda}$ of Figure~\ref{fig:toep}.
  The pattern is periodic with period $1$ and consists of a dense set
  of Bragg peaks, where increasingly smaller peaks are located at
  rational numbers whose denominators are increasingly larger powers
  of $2$. Note that only peaks corresponding to $r=0,1,2,3$ in
  Eqs.~\eqref{eq:toepFou} and \eqref{eq:toepamp} are visible here.}
\label{fig:toepdiff}
\end{figure}

Using this representation for the point set $\varLambda$, the
diffraction of the Dirac comb $\delta_{\varLambda}$ can be computed
explicitly; see Baake \& Grimm (2011b) for details. The scaled
lattices with geometrically increasing period in the union in
Eq.~\eqref{eq:pdps} give rise to Bragg peaks supported on the
corresponding dual lattices, which are successively finer integer
lattices scaled with the inverse factor. The diffraction spectrum is
pure point, and the Fourier module can be parametrised as
\begin{equation}\label{eq:toepFou}
    L^{\circledast}\, = \, \mathbb{Z}[\tfrac{1}{2}] \, = \,
    \bigl\{ \tfrac{m}{2^{r}}\mid (r=0,m\in\mathbb{Z})
    \text{ or } (r\ge 1, m\text{ odd})\bigr\}\, .
\end{equation}
The diffraction measure is of the form of Eq.~\eqref{eq:modeldiff}, with 
diffraction amplitudes
\begin{equation}\label{eq:toepamp}
    A\bigl(\tfrac{m}{2^r}\bigr) \, = \, \frac{2}{3}\, \frac{(-1)^{r}}{2^{r}}\, 
    \mathrm{e}^{2^{1-r}\pi \mathrm{i}m}
\end{equation}
at the positions in $L^{\circledast}$. The factor of $\frac{2}{3}$
reflects the density of scatterers, as two thirds of positions are
occupied.  It is no coincidence that the model set expression applies
--- in fact, the set $\varLambda$ can be described as a model set, but
with a non-Euclidean internal space; in this case, the internal space
is what is known as the space of $2$-adic integers (which essentially
consists of all fractions whose denominators are powers of $2$, but
with a specific definition of the distance of two numbers). A
schematic representation of the diffraction pattern for the period
doubling chain is shown in Figure~\ref{fig:toepdiff}.

It is easy to generalise this example to other lattice-based
substitutions in one or more dimensions; any substitution of constant
length $p$ with a coincidence in the sense of Dekking (1978), which
means that the same letter appears at the same position for the images
of all letters under a certain power of the substitution rule, is a
good candidate, because it is always pure point diffractive and
carries a natural $p$-adic structure. A well-known example of this
type is the chair tiling, in its representation as a two-dimensional
block substitution; see Robinson (1999) and Baake \& Grimm (2013) for
details.

\subsection{Order and singular continuous diffraction}

\begin{figure}[b]
\centerline{\includegraphics[width=0.8\textwidth]{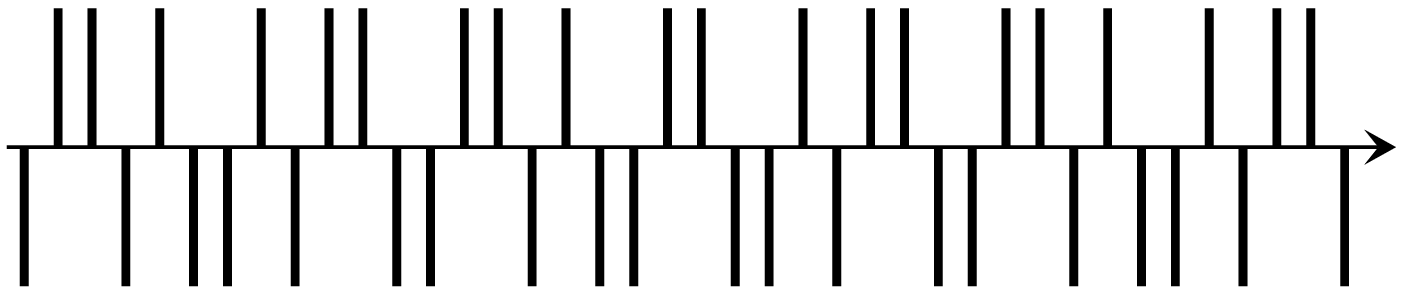}}
  \caption{Schematic representation of the Dirac comb of the
    Thue--Morse chain with weights $\pm 1$. Note that this is
    `balanced' in the sense that positive and negative weights are
    equally frequent, so the average scattering strength is zero.}
\label{fig:tmcomb}
\end{figure}

The paradigm of singular continuous diffraction is provided by the
Thue--Morse system and its generalisations (Kakutani 1972, Baake \&
Grimm 2008). Here, we consider a family of generalised Thue--Morse
substitutions (Baake, G\"{a}hler \& Grimm 2012)
\begin{equation}\label{eq:tm}
    \varrho^{(k,\ell)} : \;
    \begin{array}{r@{\;}c@{\;}l}
    1 & \mapsto & 1^{k}\,\bar{1}^{\ell} \\ 
    \bar{1} & \mapsto & \bar{1}^{k}\,1^{\ell}
    \end{array}
\end{equation}
on the two-letter alphabet $\{1,\bar{1}\}$, where
$k,\ell\in\mathbb{N}$ and the case $k=\ell=1$ corresponds to the
classic Thue--Morse case. Note that $1^j$ denotes a string of $j$
consecutive letters $1$ here.  The one-sided fixed point
$v=\varrho^{(k,\ell)}(v)$ satisfies
\begin{equation}\label{eq:tmrec}
     v_{(k+\ell)m+r} \,  = \, \begin{cases}
     v_{m}, & \text{if $0\le r\le k-1$},\\
     \bar{v}_{m}, & \text{if $k\le r\le k+\ell-1$} 
    \end{cases}
\end{equation}
where $m\ge 0$ and $0\le r\le k+\ell-1$ and $\bar{\bar{1}}=1$. It
extends (by setting $v_{-i-1}=v_{i}$ for $i\ge 0$) to a symmetric
bi-infinite fixed point word under the square of the substitution
$\varrho^{(k,\ell)}$. For instance, the symmetric bi-infinite fixed
point for the classic Thue--Morse case $k=\ell=1$ has core
\[
    \ldots \bar{1} 1 1 \bar{1}
           1 \bar{1} \bar{1} 1 
           \bar{1} 1 1 \bar{1}
           \bar{1} 1 1 \bar{1}
           1 \bar{1} \bar{1} 1 \big| 
           1 \bar{1} \bar{1} 1
           \bar{1} 1 1 \bar{1}
           \bar{1} 1 1 \bar{1}
           1 \bar{1} \bar{1} 1 
           \bar{1} 1 1 \bar{1}
    \ldots
\]
where the vertical bar denotes the origin. A schematic representation
of the corresponding Dirac comb is shown in Figure~\ref{fig:tmcomb}.

The corresponding weighted Dirac comb on $\mathbb{Z}$, interpreting
the two letters as weights (with $\bar{1}$ interpreted as $-1$),
is thus given by $\omega = \sum_{i\in\mathbb{Z}}v_{i}\delta_{i}$. Its
autocorrelation $\gamma=\sum_{m\in\mathbb{Z}}\eta(m)\delta_{m}$ is
also a Dirac comb on $\mathbb{Z}$, where the autocorrelation
coefficients $\eta(m)$ satisfy $\eta(0)=1$, $\eta(-m)=\eta(m)$ and the
recursion relation
\begin{equation}\label{eq:etarec}
\begin{split}
    \eta\bigl((k+\ell)m+r\bigr) \, = \, \frac{1}{k+\ell}
    \Bigl(& \alpha(k,\ell,r) \,\eta(m) + \\
    & \alpha(k,\ell,k +\ell - r) \,\eta(m+1)\Bigr)
\end{split}
\end{equation}
for arbitrary $m\in\mathbb{Z}$ and $0\le r\le k+\ell-1$. The recursion
follows directly from Eq.~\eqref{eq:tmrec}, with
$\alpha(k,\ell,r)=k+\ell-r-2\min(k,\ell,r,k+\ell-r)$. This system has
a unique solution, and properties of the solution show that the
corresponding diffraction measure $\widehat{\gamma}$ is purely
singular continuous; see Baake, G\"{a}hler \& Grimm (2012) for
the mathematical details of the argument.

\begin{figure}
\centerline{\includegraphics{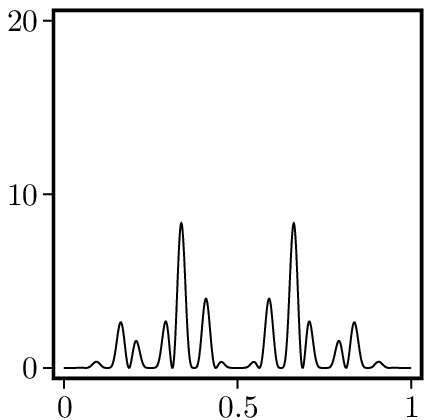}\hspace{2ex}
\includegraphics{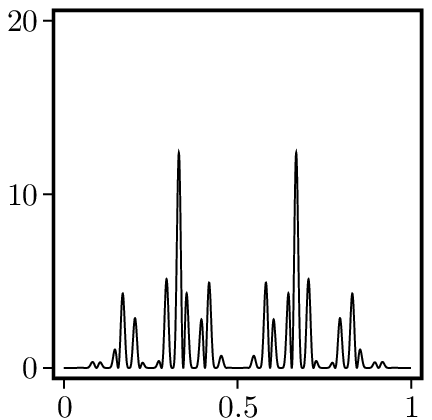}\hspace{2ex}
\includegraphics{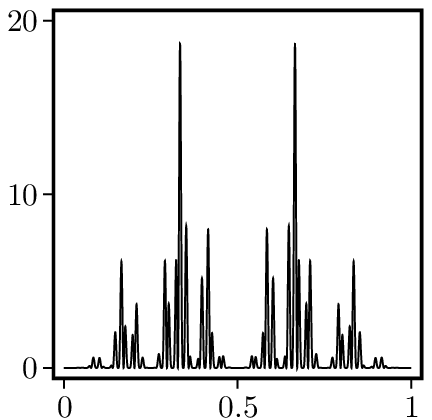}}
\centerline{\includegraphics{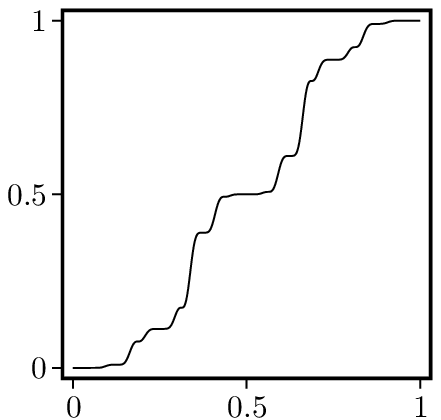}\hspace{2ex}
\includegraphics{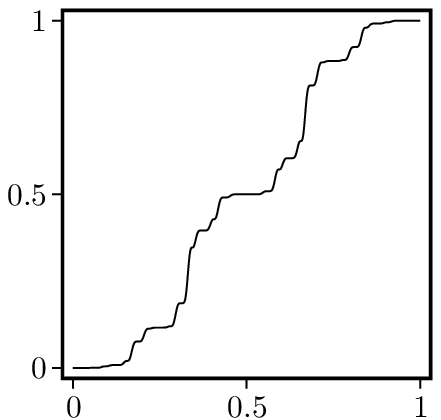}\hspace{2ex}
\includegraphics{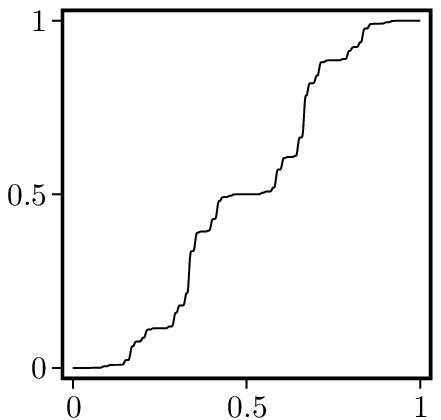}}
\caption{Approximating density functions $f^{(N)}$ (top) and 
corresponding approximating distribution function $F^{(N)}$ (bottom)
for the diffraction of the
classic Thue--Morse chain ($k=\ell=1$) with $N=4$ (left), $N=5$ (centre) and 
$N=6$ (right).}
\label{fig:tmdens}
\end{figure}

\begin{figure}
\centerline{\includegraphics{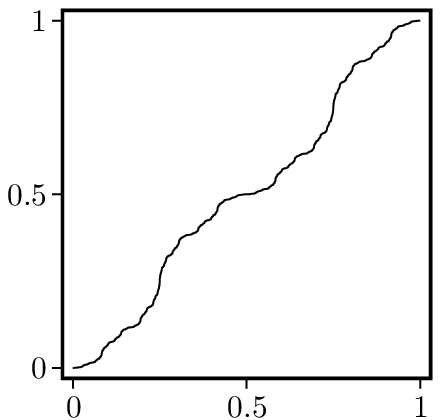}\hspace{2ex}
\includegraphics{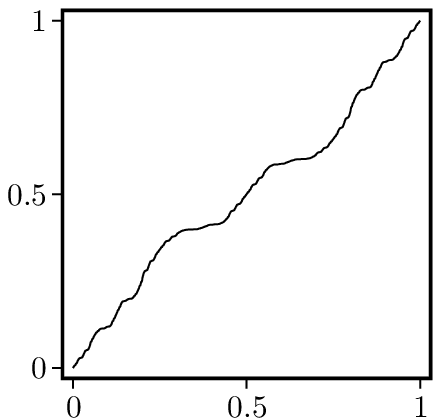}\hspace{2ex}
\includegraphics{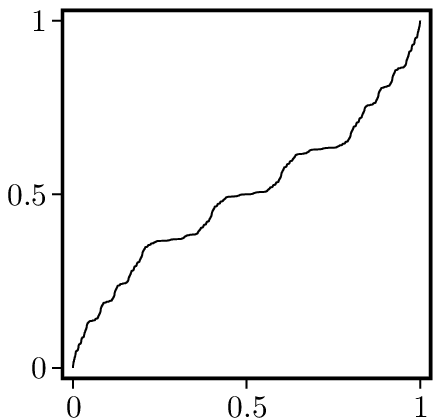}}
\caption{Continuous distribution functions for the diffraction of
  generalised Thue--Morse chains with $\ell=1$ and $k=2$ (left),
  $k=3$ (centre) and $k=4$ (right).}
\label{fig:tm}
\end{figure}

The diffraction measure is periodic with period $1$ (due to the fact
that the Dirac comb is supported on the integer lattice $\mathbb{Z}$)
and the diffraction intensity can be represented as a limit of a product,
\[
    f^{(N)}(x) \, = \prod_{n=0}^{N} 
   \vartheta\bigl((k+\ell)^{n}_{} x\bigr)\, ,
\]
which is known as a Riesz product, with
\[
    \vartheta(x) \, = \, 1+\frac{2}{k+\ell}\! 
     \sum_{r=1}^{k+\ell-1} \!\!\alpha(k,\ell,r)\, 
    \cos(2\pi rx) .
\]
The limit as $N\to\infty$ has to be considered carefully.  While the
truncated product $f^{(N)}$ is a smooth function that can be
interpreted as a density of an absolutely continuous measure, this is
not the case in the limit, because it represents a purely singular
continuous measure. Accordingly, the approximating density functions
$f^{(N)}$ become increasingly spiky with growing value of $N$; see
Figure~\ref{fig:tmdens} for an example.  Mathematically, we speak of a
limit in the vague topology. However, the corresponding distribution
function $F^{(N)}(x)=\int_{0}^{x}f^{(N)}(x)\,\mathrm{d}x$ (which
corresponds to the integrated diffraction intensity) converges and
possesses a continuous limit; compare the bottom part of
Figure~\ref{fig:tmdens}.  The limit function can be calculated and
expressed as an explicit Fourier series; several examples are shown
in Figure~\ref{fig:tm}.

While this case has no point spectrum (the trivial Bragg peak at $0$
being absent due to our balanced choice of weights, corresponding to
zero average scattering strength), it is by no means featureless. In
fact, there are peaks that grow with certain scaling exponents (in
terms of the system size) at certain points in Fourier space (the most
prominent examples can be found at rational positions $\frac{1}{3}$
and $\frac{2}{3}$ in Figure~\ref{fig:tmdens}), while the growth is not
well defined at uncountably many other positions (due to the
non-convergence of the density functions); see Baake, Grimm \& Nilsson
(2014) for a detailed account of the classic Thue--Morse case.

\begin{figure}[b]
\centerline{\includegraphics[width=0.6\textwidth]{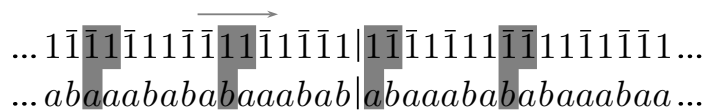}}
\caption{The action of the sliding block map $\phi$ on a Thue--Morse word
produces a period doubling word.}
\label{fig:tmtopd}
\end{figure}

Clearly, the generalised Thue--Morse systems possess hierarchical
order, although this is not reflected in a pure point component in their
diffraction measures. However, this `hidden' order is visible in other
correlations. Explicitly, it can be revealed by looking at the
two-point correlations of \emph{pairs} rather than of single points.
Looking at pairs can be described by considering the image of the
bi-infinite fixed point word $v$ under a sliding block map $\phi$,
which maps pairs of adjacent letters to $a$ or $b$ according to
whether they are equal or different, so $\phi\!:\, 11,
\bar{1}\bar{1}\mapsto a,\; 1\bar{1}, \bar{1}1\mapsto b$; see
Figure~\ref{fig:tmtopd} for an illustration.

This maps the set of generalised Thue--Morse words to bi-infinite
words which are locally indistinguishable to fixed point words of the
generalised period doubling substitution
\[
   a\mapsto b^{k-1}ab^{\ell-1}b,\quad b\mapsto b^{k-1}ab^{\ell-1}a,
\]
which reduces to the period doubling substitution (with $a=1$ and
$b=0$) in the case $k=\ell=1$.  This map is globally $2:1$, meaning
that there are precisely two generalised Thue--Morse words that are
mapped onto the same generalised period doubling word. This is most
easily seen by noticing that, when going backwards from a generalised
period doubling word, there is a single free choice for one letter $a$
or $b$ at one position, where either preimage can be chosen, after
which all other preimages are uniquely determined (due to the overlap
of adjacent pairs). As the generalised period doubling substitution
has a coincidence in the sense of Dekking (1978), it is pure point
diffractive, as discussed above for the (standard) period doubling
case. In fact, the corresponding point sets are model sets, this time
with $(k+\ell)$-adic numbers as internal space, and the pure point
diffraction is supported on the set $\mathbb{Z}[\frac{1}{k+\ell}]$,
which contains all inverse powers of $(k+\ell)$ as generating
elements.

In the language of dynamical systems, the dynamical system (where the
dynamics is given by shifting the sequence by an integer)
corresponding to generalised period doubling words constitutes a
\emph{factor} of the dynamical system associated to the generalised
Thue--Morse words. Here, the word factor refers to the fact that it is
the image under the sliding block map $\phi$. What happens in this
case is that the diffraction spectrum of the factor (the generalised
period doubling system) picks up a non-trivial point spectrum, which
is `hidden' in the Thue--Morse system, in the sense that it does not
show up in its diffraction spectrum (even in the case of general
weights). However, this pure point spectrum is part of the so-called
\emph{dynamical spectrum} of the Thue--Morse system, where the
dynamical spectrum refers to the spectrum of the operator which
generates the translation action; see Queff\'{e}lec (2010) for
details. The dynamical spectrum is, in general, richer than the
diffraction spectrum. This can be intuitively understood because
diffraction, as the Fourier transform of the autocorrelation, only
measures two-point correlations, while the dynamical spectrum `knows'
about more general properties under the shift action, so effectively
can probe higher correlations. We shall come back to this point at the
end of this section.

\subsection{Order and absolutely continuous diffraction}

Absolutely continuous (`diffuse') diffraction is commonly associated
with randomness. Indeed, stochastic systems typically show absolutely
continuous diffraction; the simplest case is the Bernoulli shift,
based on a random sequence 
\[ 
   X \, =\,  (\ldots, X^{}_{-2}, X^{}_{-1}, 
   X^{}_{0}, X^{}_{1}, X^{}_{2}, \ldots) \,\in\, \{\pm 1\}^{\mathbb{Z}}
\] 
of independent and identically distributed (i.i.d.)  events with
outcome $\pm 1$, with probability $p$ for outcome $1$ and probability
$1-p$ for outcome $-1$. The Bernoulli shift has (metric) entropy $H(p)
= - p\log (p) - (1\!-\!p) \log (1\!-\!p)$, which is greater than zero
except for the deterministic limiting cases $p=0$ and $p=1$. All the
examples discussed earlier were deterministic sequences with zero
entropy.

A random sequence $X\in\{\pm 1\}^{\mathbb{Z}}$ leads to a Dirac comb
$\omega = \sum_{j\in\mathbb{Z}} X^{}_{j} \delta^{}_{j}$, which is a
translation bounded random measure with autocorrelation
$\gamma^{}_{\mathrm{B}} =\sum_{m\in\mathbb{Z}}
\eta^{}_{\mathrm{B}}(m)\,\delta^{}_{m}$. The autocorrelation
coefficients are, almost surely (in the probabilistic sense, so with
probability $1$), given by
\[
\begin{split}
   \eta^{}_{\mathrm{B}}(m) \, = \, 
   \begin{cases}
   1,  & m = 0, \\
   (2p\! -\! 1)^2, & m\ne 0.
    \end{cases}
\end{split}
\]
as a consequence of the strong law of large numbers. The corresponding
diffraction measure then, almost surely, is given by
\[
   \widehat{\gamma^{}_{\mathrm{B}}}\, =
    \, (2p-1)^{2}\delta^{}_{\mathbb{Z}} \,+\, 
   4 p (1-p)\,\lambda\, ,
\]
which contains both pure point (for $p\ne \frac{1}{2}$) and absolutely
continuous components (for $p\not\in\{0,1\}$). The pure point part
vanishes when both weights appear with equal probability, while the
absolutely continuous part vanishes in the two deterministic, periodic
cases.

\begin{figure}
\centerline{\includegraphics[width=0.8\textwidth]{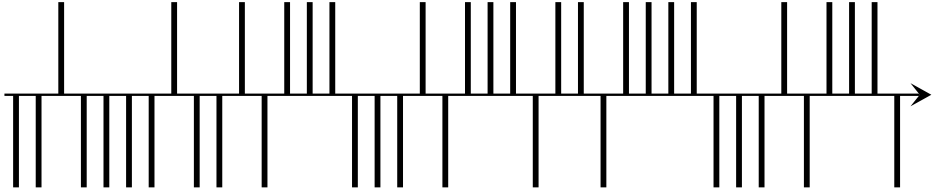}}
  \caption{Schematic representation of the Dirac comb of the
    Rudin--Shapiro chain with weights $\pm 1$, which is again
    `balanced' in the sense that the average scattering strength is
    zero.}
\label{fig:rscomb}
\end{figure}

It is, however, possible to construct deterministic systems with
absolutely continuous diffraction as well. The paradigm for this situation
is the Rudin--Shapiro chain (Rudin 1959, Shapiro 1951, Queff\'{e}lec
2010). Its binary version $w\in\{\pm 1\}^{\mathbb{Z}}$ can be defined
by initial conditions $w(-1)=-1$, $w(0)=1$, and the recursion
\begin{equation}\label{eq:rs}
   w(4n+\ell)=
    \begin{cases} w(n),  & \mbox{for $\,\ell\in\{0,1\}$,} \\
          (-1)^{n+\ell}\,w(n), & \mbox{for $\,\ell\in\{2,3\}$.}
     \end{cases}
\end{equation}
A schematic representation of the corresponding Dirac comb is shown in
Figure~\ref{fig:rscomb}.  By considering the recursion relation for
autocorrelation coefficients induced by Eq.~\eqref{eq:rs}, in a
similar way as for the generalised Thue--Morse case above, on can show
(Baake \& Grimm 2009) that the autocorrelation has the simple form
$\gamma^{}_{\mathrm{RS}} = \delta^{}_{0}$, which means that all
correlations (apart from the trivial case with distance zero) average
to zero along the chain. According to the two-point correlations, the
Rudin--Shapiro chain hence looks completely uncorrelated, exactly as
the random chain with probability $p=\frac{1}{2}$.  As a consequence,
the diffraction measure is Lebesgue measure,
$\widehat{\gamma^{}_{\mathrm{RS}}} = \lambda$, which is clearly
absolutely continuous with respect to itself. This means that the
diffraction intensity is constant in space, and hence completely
featureless, reflecting the complete absence of two-point correlation
in the structure. This example shows that two very different systems
such as the $p=\frac{1}{2}$ Bernoulli chain with entropy $\log (2)$
and the completely deterministic binary Rudin--Shapiro chain (with
zero entropy) can produce the same autocorrelation and hence the same
diffraction measure. Such structures are called homometric (Patterson
1944) and show that the inverse problem does not have a unique
solution in general.

In fact, the situation is worse than that, as from the results above
one can construct an entire one-parameter family of stochastic Dirac
combs which all are homometric with the Rudin--Shapiro chain. This is
done by the \emph{Bernoullisation} procedure (Baake \& Grimm 2009).
Applying it to the Rudin--Shapiro Dirac comb, the weight at any
position is changed randomly with probability $1-p$, resulting in
\[
    \omega \, = \, \sum_{j\in\mathbb{Z}} w^{}_{j}\, X^{}_{j}\, \delta^{}_{j} \, ,
\]
with the Rudin--Shapiro sequence $w\in\{\pm 1\}^{\mathbb{Z}}$ and the
random sequence $X\in\{\pm 1\}^{\mathbb{Z}}$ as defined above. This is
a `model of second thoughts' in the sense that, when starting from a
Rudin--Shapiro sequence, weights are randomly changed with probability
$1-p$ independently at each position along the chain.  We thus can
continuously interpolate between the Rudin--Shapiro chain with entropy
$0$ and the $p=\frac{1}{2}$ Bernoulli chain with entropy $\log (2)$,
with all systems sharing the same absolutely continuous diffraction.

It is interesting to note that the Rudin--Shapiro chain, like the
generalised Thue--Morse chains above, possesses a `hidden'
limit-periodic order that is revealed when looking at an appropriate
dynamical factor. Using the same sliding block map $\phi$ as above,
one obtains once more a factor with pure point diffraction spectrum,
in this case supported on $\mathbb{Z}[\frac{1}{2}]$, as for the period
doubling case; see Baake \& Grimm (2013) for details. Clearly, this
does not happen for the stochastic chain, which does not have any
`hidden' order.

\subsection{Discrete structures with continuous symmetry}

\begin{figure}[b]
\centerline{\includegraphics[width=0.5\textwidth]{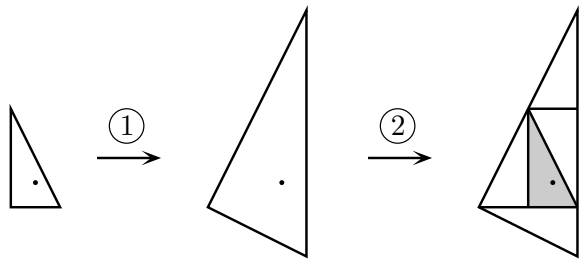}}
\caption{Inflation rule for the pinwheel tiling. The dot marks the
  reference point, and the shading emphasises that the particular
  triangle is in the original position and orientation, ensuring that
  repeated application of the inflation rule produces a fixed point
  tiling.}
\label{fig:pininf} 
\end{figure}

\begin{figure}[t]
\centerline{\includegraphics{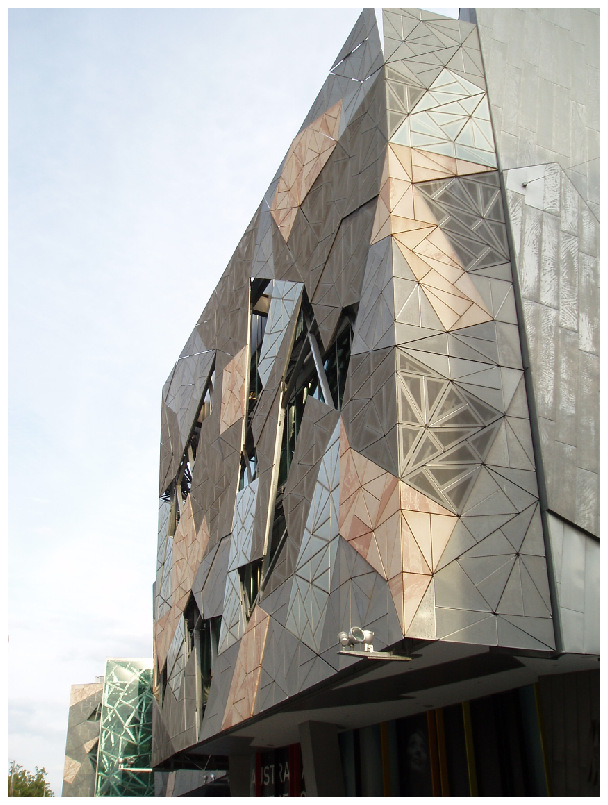}}
\caption{A building at Melbourne's Federation Square featuring
a pinwheel tiling fa\c{c}ade.}
\label{fig:melb} 
\end{figure}

An interesting (and still somewhat mysterious) class of structures is
provided by discrete systems which possess a continuous symmetry. The
paradigm for such a structure is the Conway--Radin pinwheel tiling
(Radin 1994). It is an inflation tiling based on a single triangular
prototile of edge lengths $1$, $2$ and $\sqrt{5}$ together with its
reflected version. The inflation rule is shown in
Figure~\ref{fig:pininf}; it consists of a linear rescaling by the
inflation factor $\sqrt{5}$ (first step) and the dissection of the
inflated triangle into five copies of the original prototile (second
step), where both orientations occur. The reflected rule applies to
the reflected triangle, and hence the tiling is reflection
symmetric. A realisation of the tiling is shown in
Figure~\ref{fig:melb}.

What makes this inflation rule special is the rotation it introduces
between copies of the prototiles. This rotation by an angle
$\vartheta=-\arctan(\frac{1}{2})$ is incommensurate with $\pi$, and as
a result introduces new, independent directions under inflation.
Iteration of the inflation rule on an initial patch thus leads to
patches comprising an exponentially increasing number of triangles
occurring in a linearly growing number of independent directions. In
the limit of an infinite tiling, triangles appear in infinitely many
different orientations. While for the familiar cases of Penrose-type
tilings inflation rules produce tilings with discrete (in the Penrose
case decagonal) symmetry (in the sense that the tiling space defined
by the fixed point tilings has decagonal symmetry; see Baake \& Grimm
(2013) for details), the pinwheel inflation produces a tiling space
with complete circular symmetry (Radin 1994, Radin 1997, Moody,
Postnikoff \& Strungaru 2006). As a consequence, its diffraction is
circularly symmetric as well, and hence cannot have any pure point
component apart from the trivial Bragg peak at the origin.

\begin{figure}
\centerline{\includegraphics{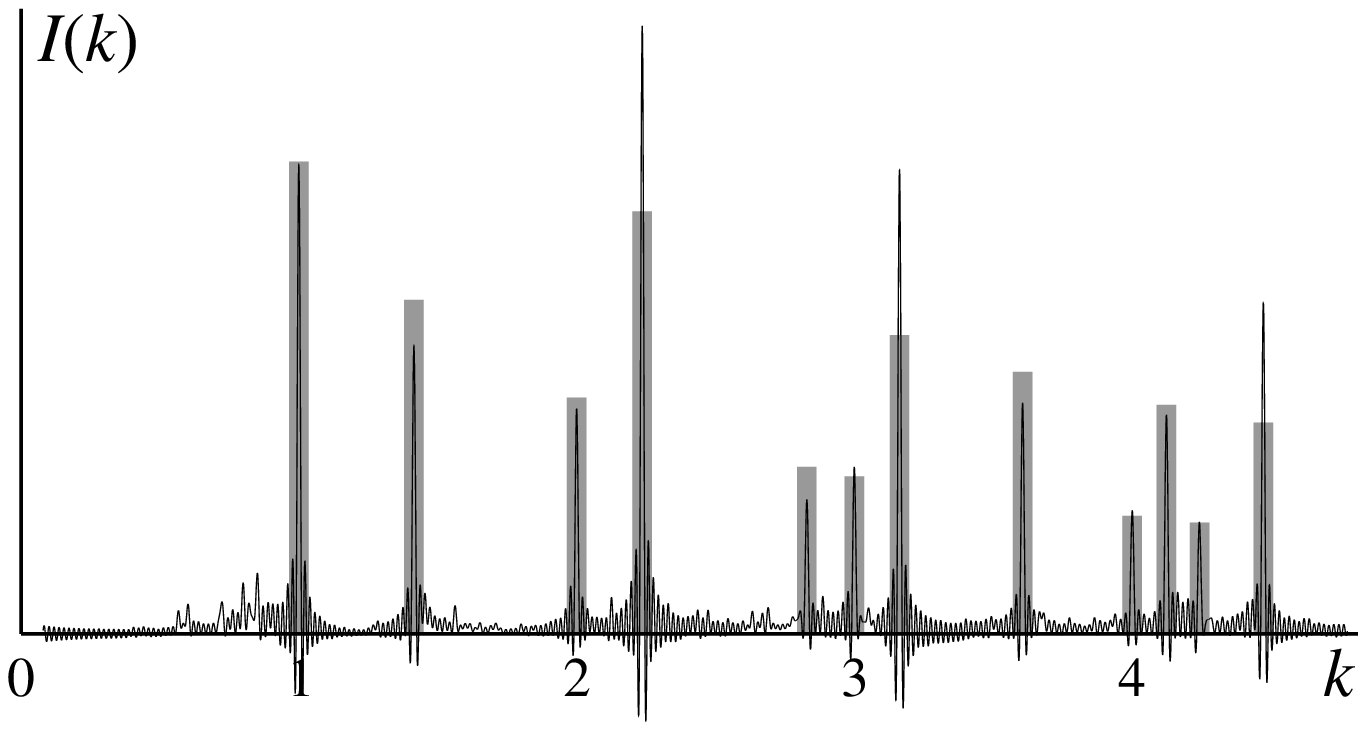}}
\caption{Approximation of radial diffraction intensity $I(k)$ for the
  pinwheel diffraction (black line), based on data from a finite
  system. The grey columns indicate the sharp rings observed in a toy
  model of powder diffraction from a planar square-lattice structure,
  with the relative scale adjusted according to the first main peak;
  see Baake, Frettl\"{o}h \& Grimm (2007a, 2007b) for details.}
\label{fig:pinsq} 
\end{figure}

In fact, the rotation is rather special, because it is a coincidence
rotation for the planar square lattice, as
$\tan(\vartheta)=-\frac{1}{2}$ is rational; see Baake (1997) for
background. This property is behind the observation that the point set
of pinwheel reference points can either be seen as a subset of rotated
square lattices or a subset of scaled square lattices, with scaling by
inverse powers of $5$ (Baake, Frettl\"{o}h \& Grimm 2007a), which
makes it possible to draw conclusions on the diffraction spectrum by
using a radial version of Poisson's summation formula. This provides
evidence that the diffraction consists of sharp rings, and that it is
surprisingly similar to a toy model of powder diffraction of square
lattice structures (Baake, Frettl\"{o}h \& Grimm 2007a, 2007b). A
diffraction measure supported on sharp rings in the plane is singular
continuous, and it is clear that the diffraction of the pinwheel
tiling contains singular continuous components of this type; however,
to date there is no complete characterisation of the diffraction
spectrum of this example. Results of numerical investigations suggest
that an absolutely continuous component may also be present. An
approximation of the radially averaged diffraction is shown in
Figure~\ref{fig:pinsq}.

While the pinwheel tiling may seem a rather exotic structure, it is
generated by a quite simple inflation rule with only a single shape up
to congruency. There are many other structures of this type; see
Frettl\"{o}h (2008) for some examples.

\subsection{Diffraction versus dynamical spectra}

The examples of the Thue--Morse and Rudin--Shapiro systems show that
systems can possess `hidden' order that does not manifest itself by a
pure point component in the diffraction pattern. However, this order
can show up in the dynamical spectrum, which is related to the
analysis of the translation action on the structure. There is a close
relationship between these two spectral quantities --- indeed, the
first proofs of the pure point diffractivity of model sets employed
the link to dynamical spectra, using the results that the diffraction
spectrum is pure point if and only if the dynamical spectrum is. In
general, however, the dynamical spectrum can be richer (van Enter \&
Mi\c{e}kisz 1992), and the Thue--Morse and Rudin--Shapiro systems are
examples; in both cases, the dynamical spectrum contains the pure
point component $\mathbb{Z}[\frac{1}{2}]$ which arises because both
examples stem from primitive substitutions of constant length $2$ (in
the Rudin--Shapiro case, the underlying substitution employs four
different letters, and the binary system is derived from this by
identifying pairs of letters; see Baake \& Grimm (2013) for details).

A particularly simple yet striking example, originally suggested by
van Enter, is discussed in Baake \& van Enter (2011). It considers the
set of certain configurations of $\pm 1$ on the integer lattice
$\mathbb{Z}$. The allowed configurations $w\in\{\pm 1\}^{\mathbb{Z}}$
are obtained by partitioning the lattice into pairs of neighbouring
points (there are two ways to do this), and then randomly assigning to
each pair either the values $(+1,-1)$ or $(-1,+1)$.  Turning a
configuration $w$ into a signed Dirac comb with weights $w_{i}\in\{\pm
1 \}$, it is easy to show, by an application of the strong law of
large numbers, that the autocorrelation is (almost surely) given by
$\gamma = \delta^{}_{0} - \frac{1}{2} (\delta^{}_{1} +
\delta^{}_{-1})$. The corresponding diffraction measure is then
\[
   \widehat{\gamma} \,=\,
    \bigl( 1 - \cos(2 \pi k) \bigr) \lambda\, ,
\]
and hence purely absolutely continuous, where the Radon-Nikodym
density relative to $\lambda$ is written as a function of
$k$. However, the dynamical spectrum of this system contains
eigenvalues (hence a pure point part), reflecting the order in the
system imposed by the `dimer' condition on pairs. This can be revealed
by considering a block map similar to to the map $\phi$ used
above. Explicitly, setting $u_{i} = - w_{i} w_{i+1}$ for
$i\in\mathbb{Z}$ maps $w$ to a new sequence $u$, which (almost surely)
has the diffraction measure
\[
     \widehat{\gamma^{}_{u}} \, = \,
     \tfrac{1}{4}\, \delta^{}_{\mathbb{Z}/2} + \tfrac{1}{2}\, \lambda\, ;
\]
see Baake \& van Enter (2011) for details. The `dimer' structure is
reflected in the presence of the pure point part supported on
$\frac{1}{2}\mathbb{Z}$, which also is the entire point part of the
dynamical spectrum.

This example again shows that the `hidden' order can also be seen in
diffraction, but not in the original system. Note that simply changing
the weights of the scatterers will not achieve this, although it may
contribute a trivial Bragg part. However, choosing a suitable factor
(or a family of factors) as an image of a continuous map such as the
sliding block map $\phi$ used above, makes it possible to detect the
`hidden' order via its diffraction. That this is a general property of
the relation between dynamical and diffraction spectrum is a recent
non-trivial insight; see Baake, Lenz \& van Enter (2013) for the
latest developments in this direction.

\section{Conclusions}

The discoveries of incommensurately modulated and aperiodically
ordered solids in the twentieth century (de Wolff 1974, Janner \&
Janssen 1977, Shechtman, Blech, Gratias \& Cahn 1984, Ishimasa, Nissen
\& Fukano 1985) have changed our view of
crystallography. Crystallography is no longer restricted to the
analysis of lattice periodic arrangements of atoms or molecules, but
takes a broader view which includes certain aperiodically ordered
structures, such as incommensurate crystals and quasicrystals. The
definition of a crystal has been amended to reflect this broader view.

The definition of a crystal is based on the currently known catalogue
of periodic and aperiodic crystals. We presently do not know of any
materials that have aperiodically ordered structures beyond
incommensurate crystals (including composite structures) and
quasicrystals. For the latter, so far only symmetries corresponding to
the smallest embedding dimension (in the sense of model sets) have
been observed, with octogonal, decagonal and dodecagonal quasicrystal
planes corresponding to projections from four-dimensional periodic
structures, and icosahedral quasicrystals being described by
projection from six-dimensional lattices. However, there is no \emph{a
  priori} reason that excludes other symmetries completely, or indeed
aperiodically ordered structures that are not described by model sets
obtained from projections of a lattice in a finite-dimensional
Euclidean space.

The definition of a crystal also reflects the current lack of
understanding of what constitutes order in matter (and more
generally), and in this sense should be seen as a working definition
that may well need to be revised in the future. In crystallography,
order is linked to diffraction, which makes sense because diffraction
is the method of choice to experimentally determine the structure of a
solid. The examples discussed above demonstrate that there are ordered
structures which are not captured by the current definition, either
because their pure point diffraction fails to be finitely generated,
or because they do not have any non-trivial point component in their
diffraction. While we do not know whether such structures are realised
in nature, it should become possible to manufacture such materials
with purpose-designed structures and properties. In this sense, these
structures are relevant and should be considered to be within
the realm of crystallography.

{}From a mathematical point of view, a more satisfying attempt at
defining order might employ the dynamical spectrum, which is a
generalisation of the diffraction spectrum. The results above are in
line with the intuition by van Enter \& Mi\c{e}kisz (1992) that an
apparent disorder at an `atomic' scale could be accompanied by order
at a `molecular' scale, with diffraction of derived factor structures
probing the latter. While diffraction itself only measures the
averaged two-point correlations in a structure, the dynamical spectrum
probes the repetitivity of a structure under translations, and hence
also higher-order correlations, which generally can distinguish
homometric systems (Gr\"{u}nbaum \& Moore 1995). While these are not
necessarily directly accessible by experiment, the additional
information contained in the dynamical spectrum is, in principle,
encoded in diffraction spectra of derived systems; see Baake, Lenz \&
van Enter (2013) for recent developments on establishing this
connection. Defining order via a non-trivial pure point component of
the dynamical spectrum would include structures such as the
Thue--Morse and Rudin--Shapiro systems, though presumably examples of
pinwheel type (for which the dynamical spectrum is not known) would be
excluded. In this sense, it probably is still not completely
satisfactory to capture all possible manifestations of order, but it
may provide a first step towards a better understanding.

In this paper, the discussion was limited to deterministic systems,
apart from the brief excursion on the Bernoulli chain. Clearly, moving
to partially ordered systems, which contain an element of stochastic
disorder, is relevant as well. Not only does even the most perfect
crystal contain some amount of disorder, but there are also
entropically stabilised structures with intrinsic configurational
disorder, among them many quasicrystalline phases. In this context,
the notion of `entropic order' is relevant, which has been
investigated in statistical physics, in particular with respect to the
physics of glasses; see, for instance, Kurchan \& Levine (2011), Sasa
(2012a, 2012b) and Wolff \& Levine (2014) for recent work along these
lines.

Although the importance of random tiling structures was pointed out
early on (Elser 1985), and while there is some good heuristic
information from scaling considerations (Henley 1999), there are as
yet very few mathematically rigorous results for non-trivial random
tiling structures in two or more dimensions, the only examples known
being related to solvable models of lattice statistical mechanics
(Baake \& H\"{o}ffe 2000).  We refer to Baake, Birkner \& Grimm (2015)
for a recent review on what is known about the diffraction of
partially ordered and stochastic systems.

\section*{Acknowledgements}

The author would like to express his gratitude to Michael Baake for
useful discussions and comments.

\section*{References}
\begin{trivlist}

\item
Authier, A. (2013).
\textit{Early Days of X-ray Crystallography},
Oxford: Oxford University Press.

\item 
Authier, A. \& Chapuis, G. (2014).
\textit{A Little Dictionary of Crystallography},
International Union of Crystallography.

\item
Baake, M. (1997).
Solution of the coincidence problem in dimensions
$d\le 4$,
in: \textit{The Mathematics of Long-Range Aperiodic Order}, 
edited by R.V.~Moody, pp.~9--44,
Kluwer: Dordrecht.

\item
Baake, M., Birkner, M. \& Grimm U. (2015).
Non-periodic systems with continuous diffraction measures,
in: \textit{Mathematics of Aperiodic Order}, edited by
J.\ Kellendonk, D.\ Lenz \& J.\ Savinien,
in press,
Boston: Birkh\"{a}user.

\item
Baake, M., Frettl\"{o}h, D. \& Grimm, U. (2007a).
A radial analogue of Poisson's summation formula with applications
to powder diffraction and pinwheel patterns.
\textit{J.\ Geom.\ Phys.} \textbf{57}, 1331--1343.
arXiv:math.SP/0610408.

\item
Baake, M., Frettl\"{o}h, D. \& Grimm, U. (2007b).
Pinwheel patterns and powder diffraction,
\textit{Philos.\ Mag.} \textbf{87}, 2831--2838.
arXiv:math-ph/0610012.

\item
Baake M., G\"{a}hler F. \& Grimm U. (2012).
Spectral and topological properties of a family of 
generalised Thue-Morse sequences,
\textit{J.\ Math.\ Phys.} \textbf{53}, 032701.
arXiv:1201.1423.

\item
Baake, M., G\"{a}hler, F. \& Grimm, U. (2013).
Examples of substitution systems and their factors.
\textit{J.\ Int.\ Seq.} \textbf{16}, article 13.2.14:\ 1--18.
arXiv:1211.5466.

\item
Baake, M. \& Grimm, U. (2007).
Homometric model sets and window covariograms.
\textit{Z.\ Krist.} \textbf{222}, 54--58.
arXiv:math.MG/0610411.

\item
Baake, M. \& Grimm, U.(2008).
The singular continuous diffraction measure of the Thue-Morse chain.
\textit{J.\ Phys.\ A:\ Math.\ Theor.} \textbf{41}, 422001:\ 1--6.
arXiv:0809.0580.

\item
Baake, M. \& Grimm, U. (2009).
Kinematic diffraction is insufficient to distinguish order from disorder.
\textit{Phys.\ Rev.\ B} \textbf{79}, 020203(R):\ 1--4 and 
\textit{Phys.\ Rev.\ B} \textbf{80}, 029903(E). arXiv:0810.5750.

\item
Baake, M. \& Grimm, U. (2011a).
Kinematic diffraction from a mathematical viewpoint.
\textit{Z.\ Krist.} \textbf{226}, 711--725.
arXiv:1105.0095.

\item
Baake, M. \& Grimm, U. (2011b).
Diffraction of limit periodic point sets,
\textit{Philos.\ Mag.} \textbf{91}, 2661--2670.
arXiv:1007.0707.

\item
Baake, M. \& Grimm, U. (2012).
Mathematical diffraction of aperiodic structures.
\textit{Chem.\ Soc.\ Rev.} \textbf{41}, 6821--6843. 
arXiv:1205.3633.

\item
Baake, M. \& Grimm, U. (2013). \emph{Aperiodic Order. Vol. 1: 
A Mathematical Invitation}. Cambridge: Cambridge University Press.

\item
Baake, M. \& Grimm, U. (2014). 
Squirals and beyond: Substitution tilings with singular continuous spectrum.
\emph{Ergodic Theory and Dynamical Systems} \textbf{34}, 1077--1102.
arXiv:1205.1384.

\item
Baake, M., Grimm, U. \& Nilsson, J. (2014).
Scaling of the Thue-Morse diffraction measure,
\textit{Acta Phys.\ Pol.\ A} \textbf{126}, 431--434.
arXiv:1311.4371

\item
Baake, M. \& H\"{o}ffe, M. (2000).
Diffraction of random tilings:\ some rigorous results, 
\textit{J.\ Stat.\ Phys.} \textbf{99}, 219--261.
arXiv:math-ph/9904005.

\item
Baake, M., Lenz, D. \&  Richard, C. (1997).
Pure point diffraction implies zero entropy for Delone sets with 
uniform cluster frequencies, 
\textit{Lett.\ Math.\ Phys.} \textbf{82}, 61--77.
arXiv:0706.1677.

\item
Baake, M., Lenz, D. \& van Enter, A.C.D. (2013).
Dynamical versus diffraction spectrum for structures with 
finite local complexity,
\textit{Preprint} arXiv:1307.5718.

\item
Baake, M. \& Moody, R.V. (2004).
Weighted Dirac combs with pure point diffraction,
\textit{J.\ reine angew.\ Math.\ (Crelle)} \textbf{573}, 61--94.
arXiv:math.MG/0203030.

\item
Baake, M., Moody, R.V. \& Schlottmann, M. (1998).
Limit-(quasi)periodic point sets as quasicrystals with
$p$-adic internal spaces,
\textit{J.\ Phys.\ A:\ Math.\ Gen.} \textbf{31}, 5755--5765.
arXiv:math-ph/9901008.

\item
Baake, M. \& van Enter, A.C.D. (2011). 
Close-packed dimers on the line:\ diffraction versus 
dynamical spectrum,
\textit{J.\ Stat.\ Phys.} \textbf{143}, 88--101.
arXiv:1011.1628.

\item
Bohr, H. (1947).
\textit{Almost Periodic Functions}, reprint,
Chelsea: New York.

\item 
Bombieri, E. \&  Taylor, J.E. (1986).
Which distributions of matter diffract? An initial investigation.
\textit{J.\ Phys.\ Colloques} \textbf{47}, C3-19--C3-28.

\item
Bragg, W.H. \& Bragg, W.L. (1913).
The reflection of X-rays by crystals.
\textit{Proc.\ Roy.\ Soc.\ A} \textbf{88}, 428--438.

\item
C\'{o}rdoba, A. (1989).
Dirac combs,
\textit{Lett.\ Math.\ Phys.} \textbf{17}, 191--196.

\item
Cowley, J.M. (1995).
\textit{Diffraction Physics}, 3rd edition,
North-Holland: Amsterdam.

\item
Dekking, F.M. (1978).
The spectrum of dynamical systems arising from substitutions of
constant length,
\textit{Z.\ Wahrscheinlichkeitsth.\ verw.\ Geb.} \textbf{41}, 221--239.

\item
de Bruijn. N.G. (1986).
Quasicrystals and their Fourier transforms.
\textit{Indag.\ Math.\ (Proc.)} \textbf{89}, 123--152.

\item
de Wolff, P.M. (1974).
The pseudo-symmetry of modulated crystal structures,
\textit{Acta Cryst.\ A} \textbf{30}, 777--785.

\item
Dworkin, S. (1993).
Spectral theory and x-ray diffraction,
\textit{J.\ Math.\ Phys.} \textbf{34}, 2965--2967.

\item
Elser, V. (1985).
Comment on ``Quasicrystals:\ A new class of
ordered structures'', 
\textit{Phys.\ Rev.\ Lett.} \textbf{54}, 1730.

\item
Frettl\"{o}h, D. (2008).
Substitution tilings with statistical circular symmetry,
\textit{Europ.\ J.\ Combin.} \textbf{29}, 1881--1893. 
arXiv:0704.2521.

\item
Friedrich, W., Knipping, P. \& von Laue, M.\ (1912).
Interferenz-Erscheinungen bei R\"{o}ntgen\-strahlen.
\textit{Sitzungsberichte der Kgl.\ Bayer.\ Akad.\ der Wiss.}, 303--322.

\item
Gr\"{u}nbaum, F.A. \& Moore, C.C. (1995).
The use of higher-order invariants in the determination of generalized
Patterson cyclotomic sets,
\textit{Acta Cryst.\ A} \textbf{51}, 310--323.

\item
Henley, C.L. (1999).
Random tiling models,
in:\ \textit{Quasicrystals:\ The State of the Art},
edited by D.~P.\ DiVincenzo \& P.~J.\ Steinhardt, 2nd edition, pp~459--560, 
World Scientific: Singapore.

\item
Hof, A. (1995)
On diffraction by aperiodic structures,
\textit{Commun.\ Math.\ Phys.} \textbf{169}, 25--43.

\item
International Union of Crystallography (1992).
Report of the Executive Committee for 1991. 
\emph{Acta Cryst. A} \textbf{48}, 922--946.

\item
Ishimasa, T., Nissen H.-U. \& Fukano, Y. (1985).
New ordered state between crystalline and amorphous in Ni-Cr
particles,
\textit{Phys.\ Rev.\ Lett.} \textbf{55}, 511--513.

\item
Janner, A. \& and Janssen, T.\ (1977).
Symmetry of periodically distorted crystals,
\textit{Phys.\ Rev.\ B} \textbf{15}, 643--658.

\item
Janssen, T. \& Janner, A. (2014). 
Aperiodic crystals and superspace concepts. 
\emph{Acta Cryst. B} \textbf{70}, 617--651.

\item
Kakutani, S. (1972). 
Strictly ergodic symbolic dynamical systems,
in: \textit{Proc.\ 6th Berkeley Symposium on Math.\ Statistics
and Probability}, edited by L.~M.\ LeCam, J.\ Neyman \&
E.~L.\ Scott, pp.\ 319--326.
Berkeley: University of California Press.

\item
Kurchan, J.\ \& Levine D.\ (2011).
Order in glassy systems,
\textit{J.\ Phys.\ A:\ Math.\ Theor.} \textbf{44}, 035001.

\item
Lenz, D. \& Moody, R.V. (2009).
Extinctions and correlations for uniformly discrete point 
processes with pure point dynamical spectra,
\textit{Commun.\ Math.\ Phys.} \textbf{289}, 907--923.
arXiv:0902.0567.

\item
Lenz, D. \& Moody, R.V. (2011).
Stationary processes with pure point diffraction.
\textit{Preprint} arXiv:1111.3617.

\item
Lenz, D. \& Strungaru, N. (2009).
Pure point spectrum for measure dynamical systems on 
locally compact Abelian groups,
\textit{J.\ Math.\ Pures Appl.} \textbf{92},  323--341.
arXiv:0704.2498.

\item
Lifshitz, R.\ (2003).
Quasicrystals: A matter of definition,
\textit{Foundatiions of Physics} \textbf{33}, 1703--1711.

\item
Lifshitz, R.\ (2007). 
What is a crystal? \textit{Z.\ Kristallogr.} \textbf{222}, 313--317.

\item
Lifshitz, R.\ (2011).
Symmetry breaking and order in the age of quasicrystals,
\textit{Isr.\ J.\ Chem.} \textbf{51}, 1156--1167.

\item
Meyer, Y. (1972).
\textit{Algebraic Numbers and Harmonic Analysis},
North Holland: Amsterdam.

\item
Moody, R.V. (2000).
Model sets:\ A survey, in:\
\textit{From Quasicrystals to More Complex Systems}, edited by
F.~Axel, F.~D\'enoyer \& J.~P.~Gazeau, pp.\ 145--166,
EDP Sciences: Les Ulis, and Springer: Berlin.
arXiv:math.MG/0002020.

\item
Moody, R.V., Postnikoff D. \& Strungaru N. (2006).
Circular symmetry of pinwheel diffraction,
\textit{Ann.\ Henri Poincar\'{e}} \textbf{7}, 711--730.

\item
Moody, R.V. \& Strungaru, N. (2004).
Point sets and dynamical systems in the autocorrelation topology,
\textit{Canad.\ Math.\ Bull.} \textbf{47}, 82--99.

\item
Mumford, D. \& Desolneux, A. (2010).
\textit{Pattern Theory: The Stochastic Analysis of Real-World Signals},
A K Peters: Natick, MA.

\item
Patterson, A.L. (1944).
Ambiguities in the X-ray analysis of crystal structures.
\textit{Phys.\ Rev.} \textbf{65}, 195--201.

\item
Queff\'{e}lec, M. (2010).
\textit{Substitution Dynamical Systems --- Spectral Analysis}, 
LNM 1294, 2nd edition, Springer: Berlin.

\item
Radin, C. (1994).
The pinwheel tilings of the plane,
\textit{Ann.\ Math.} \textbf{139}, 661--702.

\item
Radin, C. (1997).
Aperiodic tilings, ergodic theory and rotations, in:
\textit{The Mathematics of Long-Range Aperiodic Order}, 
edited by R.V.~Moody, pp.~499--519,
Kluwer: Dordrecht.

\item
Reed, M. \& Simon, B. (1980).
\textit{Methods of Modern Mathematical Physics I:\ Functional Analysis}, 
2nd edition, Academic Press: San Diego.

\item
Robinson, E.A. (1999).
On the table and the chair,
\textit{Indag.\ Math.} \textbf{10}, 581--599.

\item
Rudin, W. (1959).
Some theorems on Fourier coefficients,
\textit{Proc.\ Amer.\ Math.\ Soc.} \textbf{10}, 855--859.

\item
Sasa S.\ (2012a).
Statistical mechanics of glass transition in lattice molecule models,
\textit{J.\ Phys.\ A:\ Math.\ Theor.} \textbf{45}, 035002.

\item
Sasa S.\ (2012b).
Pure glass in finite dimensions,
\textit{Phys.\ Rev.\ Lett.} \textbf{109}, 165702.

\item
Schlottmann, M. (2000). 
Generalised model sets and dynamical systems,
in:\ \textit{Directions in Mathematical Quasicrystals},
CRM Monograph Series, vol.\ 13,
edited by M.~Baake \& R.~V.~Moody, pp.~143--159,
AMS: Providence, RI.

\item
Shapiro, H.S. (1951).
\textit{Extremal Problems for Polynomials and Power Series},
Masters Thesis, MIT: Boston.

\item
Shechtman, D., Blech, I., Gratias, D. \& Cahn, J.W. (1984).
Metallic phase with long-range orientational order and no
translational symmetry,
\textit{Phys.\ Rev.\ Lett.} \textbf{53}, 1951--1953.

\item
Steurer W. \& Deloudi S. (2009).
\textit{Crystallography of Quasicrystals:\ Concepts, Methods and 
Structures},
Springer: Berlin.

\item
Strungaru, N. (2005).
Almost periodic measures and long-range order in Meyer sets,
\textit{Discr.\ Comput.\ Geom.} \textbf{33}, 483--505.

\item
van Enter, A.C.D. \& Mi\c{e}kisz, J. (1992).
How should one define a weak crystal?
\textit{J.\ Stat.\ Phys.} \textbf{66}, 1147--1153.

\item
von Laue, M.\ (1912).
Eine quantitative Pr\"{u}fung der Theorie f\"{u}r die Interferenz-Erscheinungen 
bei R\"{o}ntgenstrahlen.
\textit{Sitzungsberichte der Kgl.\ Bayer.\ Akad.\ der Wiss.}, 363--373.

\item
Wolff, G.\ \& Levine D.\ (2014)
Ordered amorphous spin system ,
\textit{Europhys.\ Lett.} \textbf{107}, 17005.

\item 
Wolny, J., Kozakowski, B., Kuczera, P., 
Strzalka, R. \& Wnek, A. (2011).
Real space structure factor for different quasicrystals,
\textit{Israel J.\ Chem.} \textbf{51}, 1275--1291.

\end{trivlist}

\end{document}